\documentclass[12pt]{iopart}
\usepackage{graphicx}
\usepackage{color}
\usepackage{amsthm,subfigure}

\usepackage{iopams}
\begin{document}

\title[Effective quantum equations for the semiclassical Hydrogen atom]{Effective quantum equations for the semiclassical description of the Hydrogen atom}

\author{Guillermo Chac\'{o}n-Acosta}
\address{Departamento de Matem\'aticas Aplicadas y Sistemas,
Universidad Aut\'onoma Metropolitana-Cuajimalpa,  M\'exico D. F.
01120, M\'exico}\ead{gchacon@correo.cua.uam.mx}

\author{H\'ector H. Hern\'andez}
\address{Universidad Aut\'onoma de Chihuahua, Facultad de
Ingenier\'\i a, Nuevo Campus Universitario, Chihuahua 31125,
M\'exico}\ead{hhernandez@uach.mx}

\begin{abstract}
We study the Hydrogen atom as a quantum mechanical
system with a Coulomb like potential, with a semiclassical approach
based on an effective description of quantum mechanics. This
treatment allows us to describe the quantum state of the system as a
system of infinite many classical equations for expectation
values of configuration variables, their  moments and quantum dispersions. It also provides a
semiclassical description of the orbits and the evolution of
observables and spreadings and their back-reaction on the evolution.
\end{abstract}

\maketitle

\section{Introduction}\label{intro}

It is known from general principles \cite{Q} that the evolution of
the expectation value of the position operator in a general quantum
state is given by Ehrenfest's theorem, which is the quantum analogue
of Newton's second law. It can be written as $m\frac{d^2\langle
\hat{x} \rangle}{dt^2} = \left\langle
\hat{F}\left(\hat{x}\right)\right\rangle$. We notice immediately
that although the left hand side corresponds to the classical
expression, the right hand side does not, instead we have a
non-local force term which, in principle, can be interpreted statistically. When we
expand this term in Taylor series we notice that it contains
infinite quantum corrections that are expressed as powers of the
dispersions:
$$\left\langle
\hat{F}\left(\hat{x}\right)\right\rangle = F\left(\left\langle
\hat{x}\right\rangle\right) + \frac{1}{2}\Delta x^2
F''\left(\left\langle \hat{x}\right\rangle\right) + \ldots$$

The momentous quantum mechanics is based on the same idea applied to
any operator \cite{Eff}. It is similar to the low energy effective
action method which has been largely and successfully used in
quantum field theory. There, the actions of interacting theories can
be seen as quantum corrections to the classical action
\cite{EffAcc}.  Indeed, in some circumstances, the effective
equations reproduce the results of the effective action even better
than the well known WKB approximation \cite{Dias}.

The effective equations approach for quantum systems has been
developed to systematically analyze quantum effects through quantum
induced corrections to classical equations, leading to observable
phenomena as deviations from the semi-classical behavior \cite{Eff}.
These quantum corrections come from quantum backreaction effects. It
has been successfully applied to an isotropic and homogeneous model
in loop  quantum cosmology \cite{MB}, which has been used to predict
the evolution of the universe before the Big Bang \cite{MBNat}. It
has also been successfully used to analyze cosmological isotropic
models with matter \cite{MB-H} and with positive cosmological
constant \cite{D}, and also to study the effective constriction
equations for loop gravity and relativistic systems \cite{H1}.

Effective equations describe the behavior of the expectation values
in a definite state, replacing the description in terms of the
Schr\"{o}dinger equation by a system of infinite coupled equations
for both classical variables and quantum fluctuations. These
equations are particularly suitable for semi-classical states, which
may shed some light on effects that could be potentially observable.
Moreover, in those states, the system of infinite equations can be
reduced to a finite one by making some consistent truncations, as
for example in adiabatic approximation.

For some systems like the harmonic oscillator \cite{Eff} it is
possible to determine the exact properties in some state
 because the second-order moments form a
closed system and because they decouple from the classical
variables. For more complicated systems (e.g. anharmonic systems)
the equations for both classical and quantum variables are highly
coupled, so it will require additional approximations for the
equations.

In this work we use the momentous method to obtain the effective
description of a standard spinless quantum system with a Coulomb
potential, i.e. a hydrogen atom. In section \ref{one} we describe
this version of quantum mechanics with the introduction of the
moments or fluctuations as additional variables that encode the
quantum degrees of freedom. We describe the construction of the
quantum Hamiltonian, first for one dimensional systems and also for
higher dimensions.  We also compute the equations of motion for the
expectation values of basic operators for which the main ingredient
is the calculation of the Poisson algebra between quantum variables.
In section \ref{two} we compute the quantum Hamiltonian for this
system that corresponds to the classical Kepler problem with
correction terms. With this Hamiltonian we write the corresponding
effective equations up to second order in the moments, finding
twelve coupled equations: three for the classical variables and nine
for all the relevant moments. In section \ref{three}, we solve
numerically the system of effective equations and analyze the
quantum corrected behavior of the system comparing it qualitatively
to the classical case. Finally in section \ref{Discussion} we
summarize and discuss our results.

\section{Effective dynamics of quantum systems: Momentous quantum mechanics}\label{one}

Effective equations of quantum systems describe the dynamics of
expectation values of observables, as well as the evolution of its
dispersions. These equations allow us to study the quantum evolution
by analyzing how quantum effects modify classical dynamics. In
regimes where the fluctuations and dispersions are small with
respect to observables one can treat the quantum effects
perturbatively thus providing the ideal scenario to perform
numerical analysis in a semiclassical regime \cite{Eff}.

Equations of motion in effective theory  are derived from an
effective Hamiltonian, treating expectation values of observables and their
associated momenta as classical canonical variables.
Quantum fluctuations are introduced as a set of new \emph{quantum}
dynamical variables defined as follows:
\begin{equation}\label{1}
    G^{a,b} \equiv \left\langle (\hat{x}-x)^a (\hat{p}-p)^b \right\rangle_{\textrm{\tiny{Weyl}}},
    \quad a+b \geq 2
\end{equation}
where $x\equiv\langle \hat{x}\rangle $ and $p\equiv\langle
\hat{p}\rangle $, and the subscript indicates that the operators
inside the
brackets are Weyl (or completely symmetrical) ordered. 
This quantum variables describe the spreading of the quantum
modified evolution from the classical one.

The momenta are not arbitrary but subject to generalized uncertainty
relations such as
\begin{equation}\label{2}
    G^{2,0}G^{0,2} - (G^{1,1})^2 \geq \frac{\hbar^2}{4}.
\end{equation}
Notice that $G^{2,0}$ and $G^{0,2}$ are the standard dispersions
$\Delta x^2$ and $\Delta p^2$, and that (\ref{2}) simplifies to the
usual uncertainty principle for pure states \cite{Q}.

Evolution is obtained by evaluating Poisson brackets of variables
and momenta with the \emph{quantum} effective Hamiltonian that is
defined as the expectation value of the standard Hamiltonian
operator
\begin{equation}\label{3}
    \langle \hat{H} \rangle    \equiv H_{Q} = H(x,p) + \sum_{a,b} \frac{1}{a!b!} \frac{\partial^{a+b} H}{\partial x^a \partial
    p^b} G^{a,b},
\end{equation}
where $H(x,p)$ is the classical Hamiltonian. Equations of motion are
obtained in the usual Hamiltonian formulation $\dot{f} = \{f,H_Q\}$.
Quantum variables $G^{a,b}$ are now dynamical, as the classical ones
($x, p$). For general models one obtains an infinitely coupled
system of equations with infinitely many variables that, however
complicated, provides us a full description of the system. It is
also important to note that the effective equations so obtained are
state dependent, since the dynamic quantum fluctuations affect the
behavior of the expectation values.

To obtain effective equations of motion one compute the Poisson
brackets of variables with the quantum-corrected Hamiltonian: the
following relations, for one degree of freedom, are useful
\begin{eqnarray}
  \{x,p\} &=& 1, \label{4.1}\\
  \{x,G^{a,b}\} &=& 0, \label{4.2}\\
  \{p,G^{a,b}\} &=& 0. \label{4.3}
\end{eqnarray}
The Poisson algebra for the moments was originally obtained in
\cite{Eff} but was recently reexamined in \cite{D} for its algebraic
and numerical implementation
\begin{eqnarray}\label{5}
    \{G^{a,b},G^{c,d}\} &=& ad G^{a-1,b} G^{c,d-1} - b c G^{a,b-1}
    G^{c-1,d} + \nonumber \\
    &+& \sum_{n} \left(\frac{i \hbar}{2}\right)^{n-1}
    K^n_{abcd} G^{a+c-n,b+d-n},
\end{eqnarray}
where the sum runs over odd numbers from $n=1 \ldots \tilde{N} $,
with $1 \leq \tilde{N} < \min[a+c,b+d,a+b,c+d]$, and the coefficient
is
\begin{equation}\label{6}
    K^n_{abcd} = \sum_{s=0}^n (-1)^s s!(n-s)! \left(
                                                \begin{array}{c}
                                                  a \\
                                                  s \\
                                                \end{array}
                                              \right) \left(
                                                        \begin{array}{c}
                                                          b \\
                                                          n-s \\
                                                        \end{array}
                                                      \right) \left(
                                                                \begin{array}{c}
                                                                  c \\
                                                                  n-s \\
                                                                \end{array}
                                                              \right)\left(
                                                                       \begin{array}{c}
                                                                         d \\
                                                                         s \\
                                                                       \end{array}
                                                                     \right).
\end{equation}

For the  quantum harmonic oscillator it
was shown that the ground state energy is added to the classical
Hamiltonian \cite{Eff}, \cite{LQC-dy}. It was also seen that the
system is solvable since the moments are not coupled with the
expectation values of classical variables: there is no quantum
back-reaction. In the case of anharmonic systems, one can use an
adiabatic approximation that produces effective forces coming from
the coupling terms between the expectation variables and the
moments. This method has also been applied to the isotropic and
homogeneous model in loop quantum cosmology \cite{MB, MBNat}, and to
study different cosmological models with matter \cite{MB-H},
cosmological constant \cite{D}, among others.

For the case of $k$ pairs of canonical degrees of freedom we have
the general definition for quantum variables:
\begin{eqnarray}\label{1k}
    \fl G^{a_1,b_1,\ldots,a_k,\ldots,b_k} \equiv \left\langle  (\hat{x}_1-x_1)^{a_1} (\hat{p}_1-p_1)^{b_1}
     \cdots (\hat{x}_k-x_k)^{a_k} (\hat{p}_k-p_k)^{b_k}
    \right\rangle_{\textrm{\tiny{Weyl}}}
\end{eqnarray}
where $ a_i+b_i \geq 2$ and $i = 1,\ldots, k$. We have
$G^{0,0,\ldots,0,0}=1$,  $G^{0, \ldots,a_j, \ldots,0}=0$. We may
also note that we will always have an even number of indices and
each pair $(a_i,b_i)$ corresponds to the moments of each canonical
pairs $(x_i,p_i)$.

The generalized quantum-corrected effective Hamiltonian is as
follows
\begin{eqnarray} \label{Hamilton-severaldof}
  \fl H_{Q} := \sum_{a_1,b_1}^{\infty} \cdots \sum_{a_kb_k}^{\infty} \frac{1}{a_1!b_1! \cdots a_k!b_k!}\,\, \frac{\partial^{a_1+b_1+ \ldots + a_k+b_k}
  \ H}{\partial x_1^{a_1}\partial p_1^{b_1}\ldots\partial x_k^{a_k}\partial p_k^{b_k}}  \ G^{a_1,b_1; \ldots; a_k,
  b_k}
 \end{eqnarray}
 With this Hamiltonian we will obtain
the evolution equations for each degree of freedom. For this we need
the set of Poisson brackets: those among the expectation values are
known and those between the expectation values and moments vanish.

The Poisson algebra among the moments is the generalization of
equation (\ref{5}); this expression was first found in \cite{Eff}
and then corrected in \cite{D}
\begin{eqnarray}\label{5k}
  \fl \{G^{a_1,b_1;\ldots;a_k,b_k},G^{c_1,d_1;\ldots;c_k,d_k}\} = \sum_{i=1}^k \left( a_id_i G^{a_1,b_1;\ldots;a_i-1,b_i;\ldots;a_k,b_k} G^{c_1,d_1;\ldots,c_i,d_i-1;\ldots;c_k,d_k}
     - \right. \nonumber\\
    \left. b_i c_i G^{a_1,b_1;\ldots;a_i,b_i-1;\ldots;a_k,b_k}G^{c_1,d_1;\ldots;c_i-1,d_i;\ldots;c_k,d_k}
   \right) \nonumber\\
    +\sum_n \sum_s  \sum_{e_1,\ldots, e_k} (-1)^s \left(\frac{i \hbar}{2}\right)^{n-1}\delta_{\sum_ie_i,n}
    \nonumber\\
    \times \ \mathcal{K}^{n,s,\{e\}}_{\{a\},\{b\},\{c\},\{d\}}\, G^{a_1+c_1-e_1,b_1+d_1-e_1; \ldots; a_k+c_k-e_k,b_k+d_k-e_k} \nonumber \\
\end{eqnarray}

$\, n = 1, \ldots, \tilde{N} $, and
\begin{eqnarray}\label{enhe}
    \fl \tilde{N} = \left\{
                  \begin{array}{ll}
                    1, & \hbox{$\sum_i (\min[a_i, d_i] + \min[b_i, c_i])\leq 1$,} \\
                    \sum_i (\min[a_i, d_i] + \min[b_i, c_i])-1, & \hbox{$\sum_i (\min[a_i, d_i] + \min[b_i, c_i])>1$}.
                  \end{array}
                \right.
\end{eqnarray}

$ s = 0,\ldots, n\,$; $\,0\leq e_i\leq \min[a_i, d_i, s] + \min[b_i,
c_i, n - s]$.

The $\mathcal{K}$ coefficient  is
\begin{eqnarray}\label{6k}
    \fl \mathcal{K}^{n,s,\{e\}}_{\{a\},\{b\},\{c\},\{d\}} = \sum_{g_1,\ldots,g_k}
    \frac{\delta_{\sum_ig_i,n-s}}{s!(n-s)!}\prod_i \frac{\left(
                                                           \begin{array}{c}
                                                             a_i \\
                                                             e_i-g_i \\
                                                           \end{array}
                                                         \right)\left(
                                                                  \begin{array}{c}
                                                                    b_i \\
                                                                    g_i \\
                                                                  \end{array}
                                                                \right)\left(
                                                                         \begin{array}{c}
                                                                           c_i \\
                                                                           g_i \\
                                                                         \end{array}
                                                                       \right)\left(
                                                                                \begin{array}{c}
                                                                                  d_i \\
                                                                                  e_i-g_i \\
                                                                                \end{array}
                                                                              \right)
    }{\left(
        \begin{array}{c}
          n-s \\
          g_i \\
        \end{array}
      \right)\left(
               \begin{array}{c}
                 s \\
                 e_i-g_i \\
               \end{array}
             \right)
    },
\end{eqnarray}
where $\,\max[e_i - s, e_i - a_i, e_i - d_i, 0]\leq g_i \leq
\min[b_i, c_i, n - s, e_i]$.

Henceforth we will consider the case $k = 2$, corresponding to a two
dimensional problem, being the case for the Kepler system.

\section{Effective quantum Hamiltonian for Coulomb potential}\label{two}

In non-relativistic quantum mechanics
 the hydrogen atom is studied with the Schr\"odinger equation
with a central potential; when one considers stationary solutions
it is possible to separate the angular and the
radial parts and thus obtaining the corresponding eigenfunctions. It
is well known that for the Coulomb potential $V(r)
= - \frac{k}{r}$, with $k = \frac{e^2}{4 \pi \epsilon_0}$, the
radial function is the product of the associated Laguerre
polynomials by a decaying exponential of $r$ and $r^l$ \cite{Q}.
Moreover, the corresponding energy levels of the Hydrogen atom
are $E_n = -
\frac{k}{2 a_0 n^2}$ in
terms of the Bohr radius $a_0= \frac{\hbar^2}{m k}$.

The Schr\"odinger treatment of the Hydrogen atom has long been known
and solved \cite{Dirac}. The state of any quantum mechanical system
in this picture is encoded in the wave function
$\psi(\mathbf{r},t)$, observables and expectation values are all
obtained from the wave function considering it as the probability
distribution for the system at hand.

We study now this system using the effective momentous method
of quantum mechanics exposed in the previous section. We will
analyze how the moments evolve in time and how they modify the
corresponding classical dynamics.

The classical Hamiltonian for the Hydrogen atom is that of the
Kepler problem
\begin{equation}\label{9}
    H(r,p) = \frac{p_r^2}{2m} + \frac{p_{\theta}^2}{2mr^2} - \frac{k}{r},
\end{equation}
which, by virtue of the conservation of angular momentum, is a two
dimensional system with classical polar configuration variables $(r,
\theta)$ and
 $(p_r, p_{\theta} )$ are their canonical conjugate momenta.

Using the definition (\ref{1k}) of the moments for our two pairs of
canonical variables we have
 \begin{equation} \label{Gs-kepler}
 G^{a,b,c,d} = \left\langle  (\hat{r}-r)^{a} (\hat{p}_r-p_r)^{b} (\hat{\theta}-\theta)^c (\hat{p}_{\theta}-p_{\theta})^d
    \right\rangle_{\textrm{\tiny{Weyl}}},
 \end{equation}
from (\ref{Hamilton-severaldof}) we obtain the corresponding quantum
effective Hamiltonian
\begin{eqnarray}
  \fl H_Q = \frac{p^2_r}{2m} + \frac{p_{\theta}^2}{2mr^2} - \frac{k}{r} + \frac{G^{0,2,0,0}}{2m} + \frac{G^{0,0,0,2}}{2mr^2}  +  \sum_{a\geq 2} \left[\frac{p_{\theta}^2(a+1)}{2mr} - k \right]\frac{(-1)^a}{r^{a+1}}
    G^{a,0,0,0} \nonumber\\
    + \sum_{b\geq 1}\frac{(-1)^b}{m r^{b+2}}(b+1)\left[ p_{\theta}G^{b,0,0,1} + \frac{1}{2}G^{b,0,0,2} \right]. \label{10}
\end{eqnarray}
The first three terms correspond to the classical Hamiltonian
(\ref{9}), the fourth term is the radial momentum dispersion, the
fifth term is the angular momentum dispersion. The sum for
$a\geq 2$ is associated with the fluctuations in the radial
component, while the sum for $b\geq 1$ contains the
coupled moments for the fluctuations in $r$ and $p$.

As can be seen from (\ref{10}) the system is highly coupled, in
contrast with the case of the harmonic oscillator where the moments
decouple from the expectation values and its contribution to the
quantum Hamiltonian is constant, rendering the ground state energy
$\frac{\hbar \omega}{2}$ \cite{Eff}. The fact that for the central
potential the moments can not be decoupled may hinder the
calculation, however we will perform consistent truncations at
different orders in quantum variables. We note
 that, to second order in the moments, the last term
of the second sum does not appear, that is, in order to consider all the terms of
the effective Hamiltonian to this order we need to consider at least the
third order momenta.

We obtain the equations of motion for this set of variables by
computing their Poisson brackets with the Hamiltonian (\ref{10}).
For expectation values
\begin{equation}\label{11}
    \dot{r} = \{r,H_Q\} = \frac{p_r}{m},
\end{equation}
\[\dot{p_r} = \{p_r,H_Q\} = \frac{p_{\theta}^2}{mr^3}  - \frac{k}{r^2}   + \frac{G^{0,0,0,2}}{mr^3} + \sum_{a\geq 2} (-1)^a \frac{(a+1)}{r^{a+2}}G^{a,0,0,0}\left[ \frac{p_{\theta}^2}{2m}\frac{(a+2)}{r} - k
     \right]\]
\begin{equation}\label{12}
     + \sum_{b\geq 1} (-1)^b
     \frac{(b+1)(b+2)}{mr^{b+3}}\left[p_{\theta}
     G^{b,0,0,1}+ \frac{1}{2} G^{b,0,0,2}\right],
\end{equation}
\begin{equation}\label{11bis}
    \dot{\theta} = \{\theta,H_Q\} = \frac{p_{\theta}}{mr^2} - \frac{2}{mr^3}G^{1,0,0,1} + \sum_{a\geq 2}
    \frac{(-1)^a(a+1)}{mr^{a+2}}\left[p_{\theta}G^{a,0,0,0} + G^{a,0,0,1}\right],
\end{equation}
\begin{equation}\label{12bis}
    \dot{p_{\theta}} = \{p_{\theta},H_Q\} = 0.
\end{equation}
Equation (\ref{11}) is just the usual definition of the momentum
associated with $r$ while equation (\ref{12bis}) states the
classical conservation of angular momentum, i.e. $p_{\theta}= l =
const$. One can see the strong quantum back-reaction of this
semiclassical approach for classical variables in equations
(\ref{12}) and (\ref{11bis}).

The equations of motion for moments at any order follow from
(\ref{5k}):
\begin{eqnarray}
  \fl \dot{G}^{a_1,b_1,a_2,b_2} = \{G^{a_1,b_1,a_2,b_2},H_Q\} = \frac{1}{2m}
  \{G^{a_1,b_1,a_2,b_2},G^{0,2,0,0}\} + \frac{1}{2mr^2}
  \{G^{a_1,b_1,a_2,b_2},G^{0,0,0,2}\} \nonumber\\
  + \sum_{a\geq 2} \left[\frac{p_{\theta}^2(a+1)}{2mr} - k
  \right]\frac{(-1)^a}{r^{a+1}}\{G^{a_1,b_1,a_2,b_2},G^{a,0,0,0}\}
  \nonumber \\
  + \sum_{b\geq 1}\frac{(-1)^b}{m r^{b+2}}(b+1)\left[ p_{\theta}\{G^{a_1,b_1,a_2,b_2},G^{b,0,0,1}\} +
  \frac{1}{2}\{G^{a_1,b_1,a_2,b_2},G^{b,0,0,2}\}
  \right] \nonumber \\
  \label{13}
\end{eqnarray}
As a simplification we consider an expansion in equations
(\ref{10})-(\ref{13}) up to second order in quantum variables. The
corresponding system is
\begin{eqnarray}
  \fl \dot{r} = \frac{p_r}{m}, \label{rdot} \\
  \fl \dot{p_r} = \frac{l^2}{mr^3}  - \frac{k}{r^2}   + \frac{G^{0,0,0,2}}{mr^3} +  \frac{3}{r^{4}}G^{2,0,0,0}\left[ \frac{l^2}{m}\frac{2}{r} - k \right]
  - \frac{6l}{mr^{4}} G^{1,0,0,1},\\
  \fl \dot{\theta} = \frac{l}{mr^2} - \frac{2}{mr^3}G^{1,0,0,1} +
    \frac{3l}{mr^{4}}G^{2,0,0,0}, \label{theta} \\
  \fl \dot{G}^{1,1,0,0} = -\frac{1}{m} G^{0,2,0,0} + \left[\frac{3l^2}{2mr} - k
  \right]\frac{2}{r^{3}}G^{2,0,0,0} - \frac{2l}{m r^{3}}G^{1, 0, 0, 1}, \\
  \fl \dot{G}^{2,0,0,0} = -\frac{2}{m} G^{1,1,0,0}, \\
  \fl \dot{G}^{0,2,0,0} =  4\left[\frac{3l^2}{2mr} - k
  \right]\frac{1}{r^{3}}G^{1,1,0,0} - \frac{4l}{m r^{3}} G^{0, 1, 0, 1}, \\
  \fl \dot{G}^{0,0,1,1} =  -\frac{1}{mr^2} G^{0, 0, 0, 2}  + \frac{2l}{m r^{3}}G^{1, 0, 0, 1}, \\
  \fl \dot{G}^{0,0,2,0} = -\frac{2}{mr^2} G^{0,0,1,1}  + \frac{4l}{m r^{3}} G^{1, 0, 1, 0}, \\
  \fl \dot{G}^{0,0,0,2} = 0, \label{0002-O2} \\
  \fl \dot{G}^{1,0,1,0} = -\frac{1}{m} G^{0, 1, 1, 0} - \frac{1}{mr^2} G^{1, 0, 0, 1} + \frac{2l}{m r^{3}}G^{2, 0, 0, 0}, \\
  \fl \dot{G}^{1,0,0,1} = -\frac{1}{m}  G^{0, 1, 0, 1}, \label{g1001} \\
  \fl \dot{G}^{0,1,0,1} =  \left[\frac{3l^2}{2mr} - k
  \right]\frac{2}{r^{3}} G^{1, 0, 0, 1} - \frac{2l}{m r^{3}}G^{0, 0, 0, 2},  \label{0101} \\
  \fl \dot{G}^{0,1,1,0} = -\frac{1}{mr^2}  G^{0, 1, 0, 1} + \left[\frac{3l^2}{2mr} - k
  \right]\frac{2}{r^{3}} G^{1, 0, 1, 0} + \frac{2l}{m r^{3}}\left(G^{1,1,0,0}-G^{0,0,1,1}\right),\nonumber \\
  \label{gdot}
\end{eqnarray}
where we already inserted $p_{\theta}=l=const.$.

One can see that if all the moments $G^{a,b,c,d}$ are set to zero,
we recover the classical equations of motion. Equation
(\ref{0002-O2}) can immediately be solved $G^{0,0,0,2} = \Delta l^2
\equiv \textrm{const}$. Indeed, from (\ref{13}) it can be seen that
if $a_1=b_1=a_2=0$, $b_2=n$, then $\dot{G}^{0,0,0,b_2}$ is always
zero and all the dispersions of the angular momentum are constants,
$\Delta l^n = const$. With this simplification the system reduces to
a set of twelve coupled differential equations for classical and
quantum variables. It is evident that we will not find non-spreading
solutions as in the harmonic oscillator case \cite{Eff,LQC-dy}.

\section{Evolution and quantum back-reaction}\label{three}

In this section we analyze the evolution and quantum-corrected
behavior of the system under consideration according to equations
(\ref{rdot})-(\ref{gdot}). We discuss two cases of interest.

\subsection{Zero angular momentum}

The case $l=0$ corresponds to the so called one dimensional Hydrogen
atom. Although one might think that this system as physically
uninteresting it actually has interesting features \cite{1DH}. It
has been used to model Hydrogen atoms in the presence of strong
magnetic fields as in astrophysical systems; it is also useful in
modeling Rydberg atoms in external fields, the behavior of certain
electrons near the surface of helium \cite{1DHaps}.

\begin{figure}[h!]
\centering \subfigure[]{\includegraphics[width=6cm]{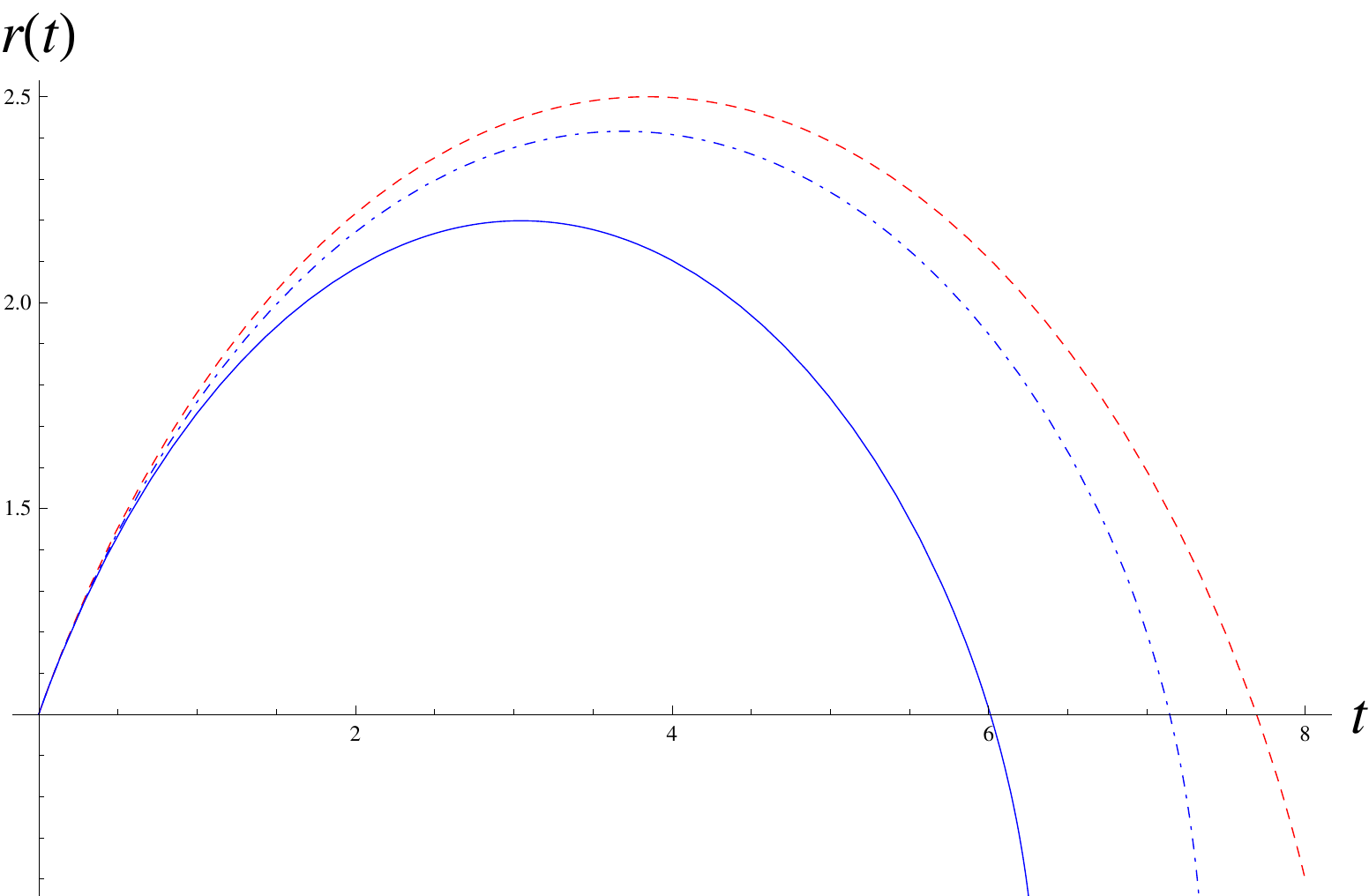}}\hfill
\subfigure[]{\includegraphics[width=6cm]{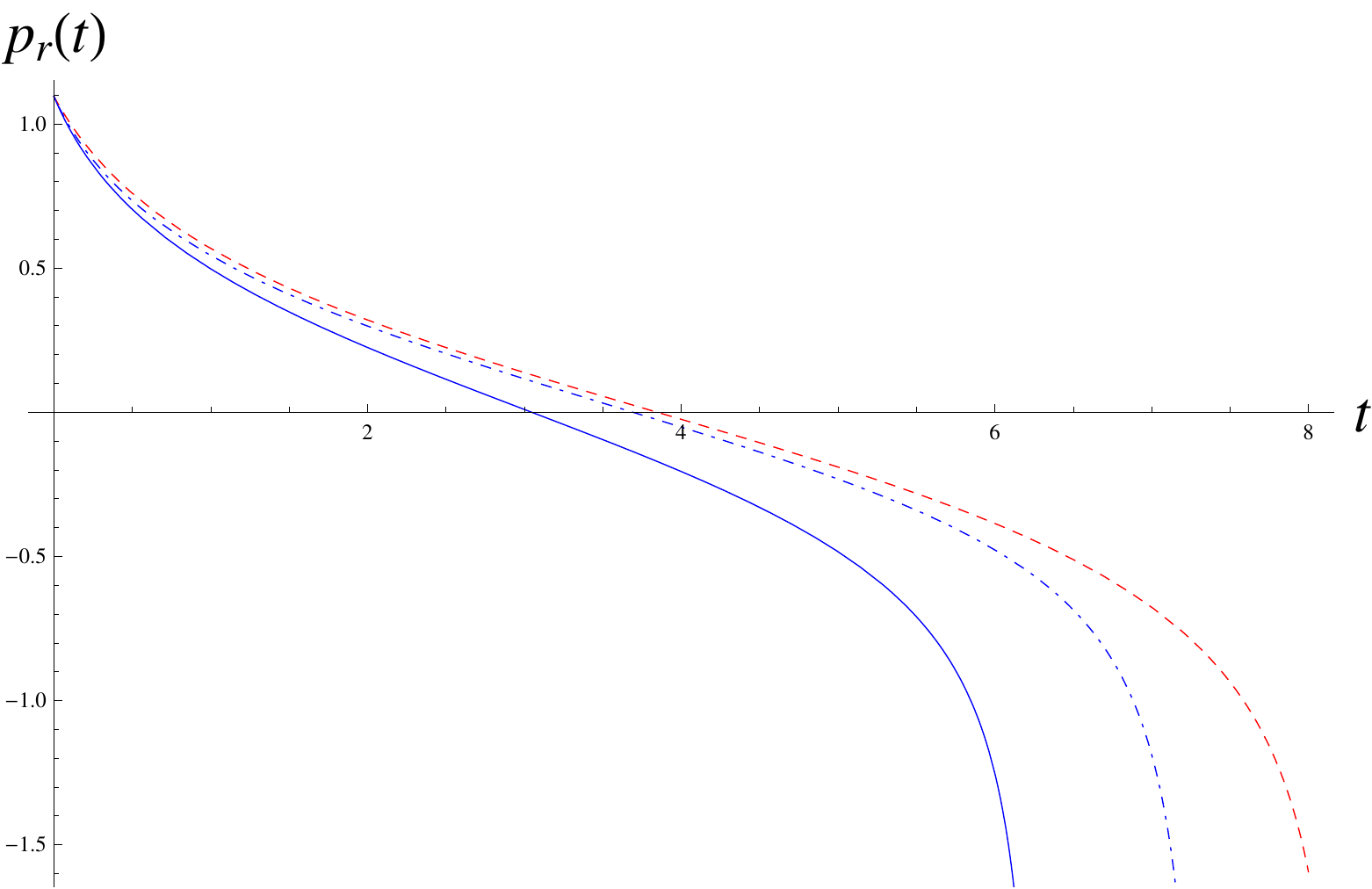}}\\
\caption[]{(a) The position $r(t)$  and (b) the momentum $p_r(t)$ as
functions of time, for $l=0$. The dashed (red) line corresponds to
the classical behavior for $m=k=1$. The other curves correspond to
the second order effective quantum behavior; when $\Delta l^2 = 0$,
solid (blue) curve and for $\Delta l^2 \neq 0$ dot-dashed (blue)
curve.} \label{r-p-l0}
\end{figure}

\begin{figure}[h!]
\centering
\includegraphics[height=6.4cm]{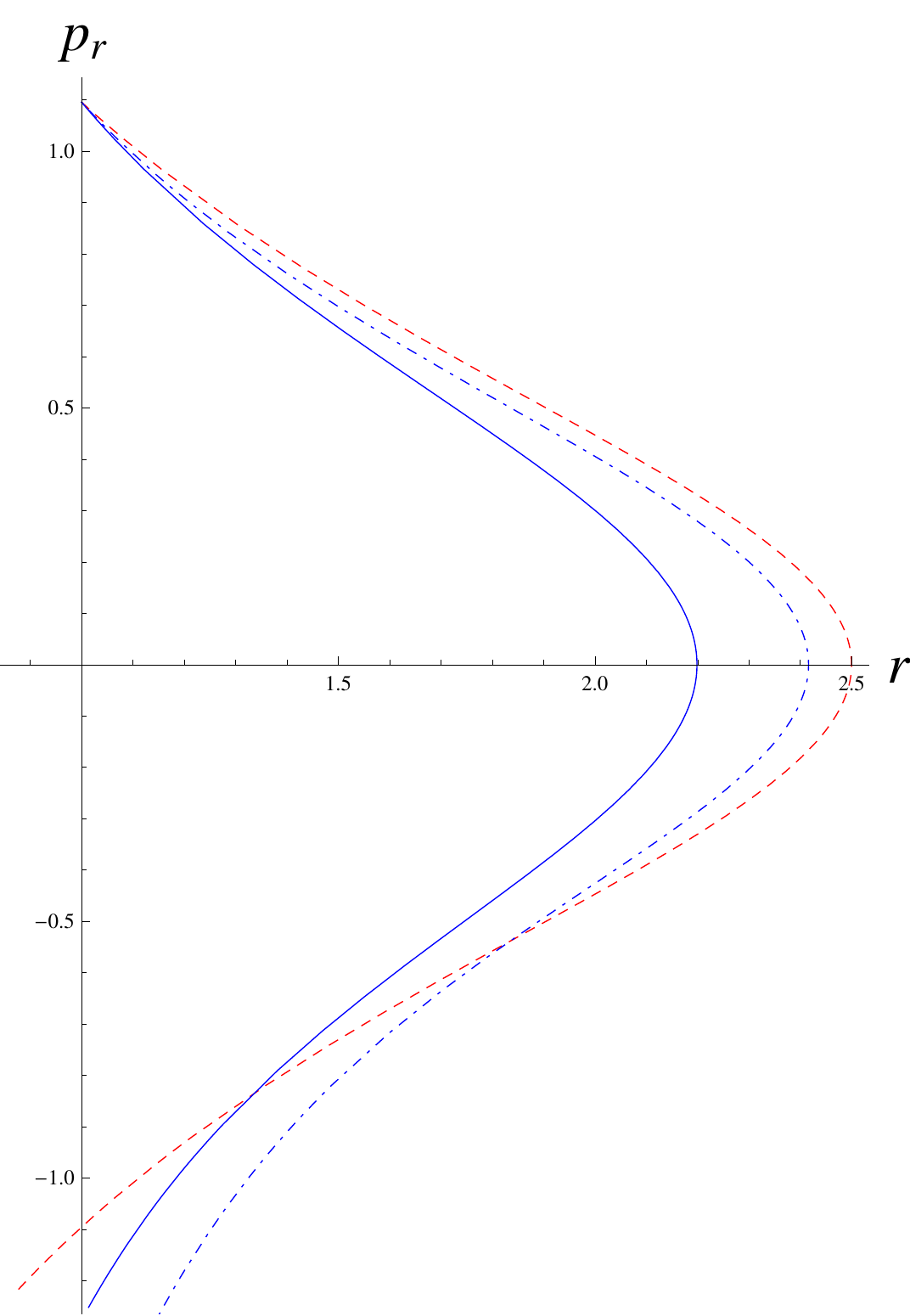}
\caption{Classical and quantum corrected phase space for $l=0$.  The
dashed (red) is the classical behavior, the quantum evolution for
$\Delta l^2 = 0$, solid (blue) curve and for $\Delta l^2 \neq 0$
dot-dashed (blue) curve.} \label{plot1}
\end{figure}

Classically this case reduces to a one-dimensional problem with
radial equations:
\begin{equation}\label{rclass_l}
    \dot{r} = \frac{p_r}{m},\qquad \dot{p_r} = -\frac{k}{r^2}.
\end{equation}
As it evolves in time the position of the particle begins to
increase as its velocity decreases, after that the radius
starts to decrease as the velocity increases in the opposite
direction (see Fig. \ref{r-p-l0}). In Fig. \ref{plot1}, we show the
phase space of the system in which these features are clear.

\begin{figure}[h!]
\centering \subfigure[]{\includegraphics[width=6cm]{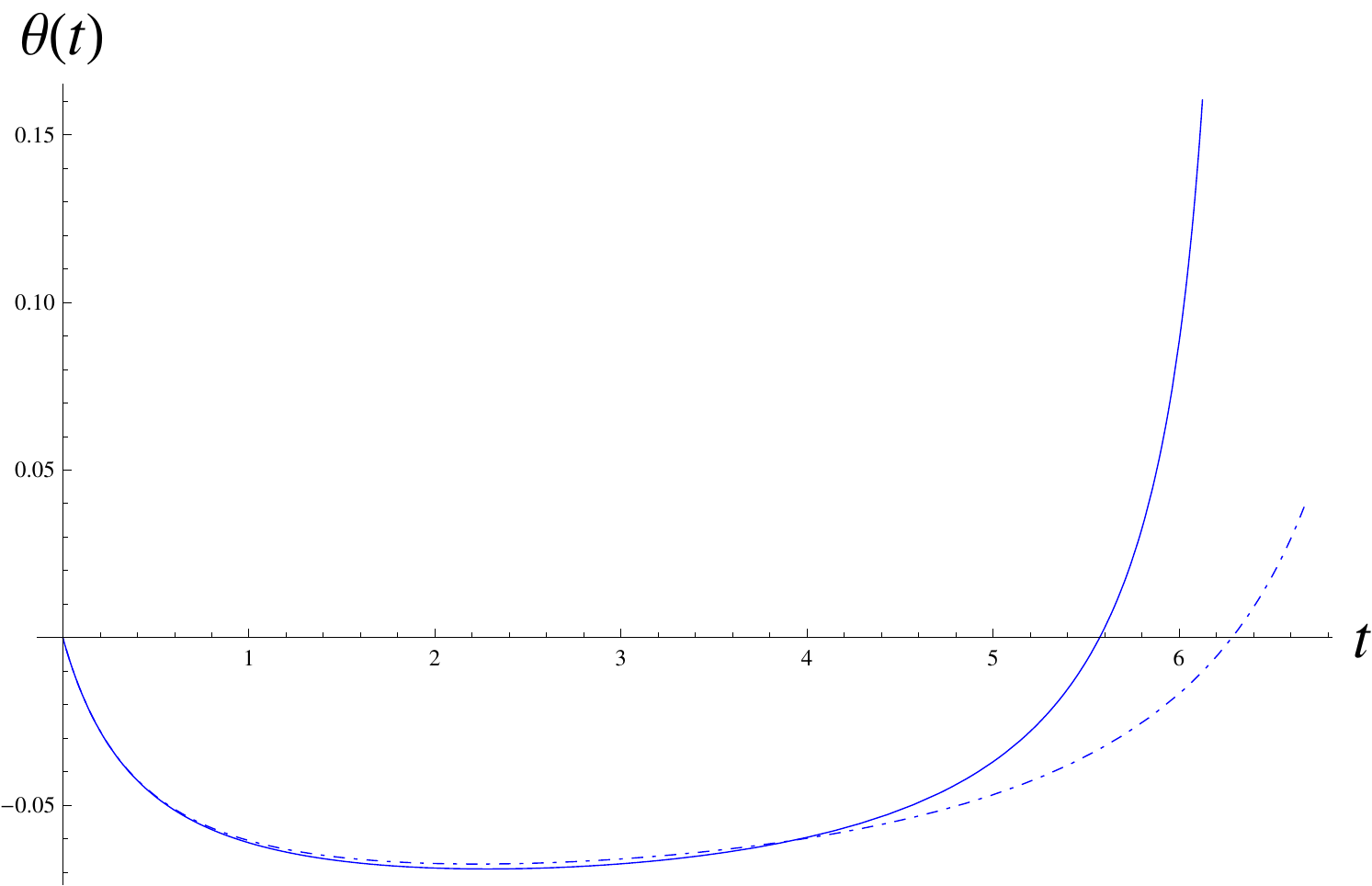}}\hfill
\subfigure[]{\includegraphics[width=6cm]{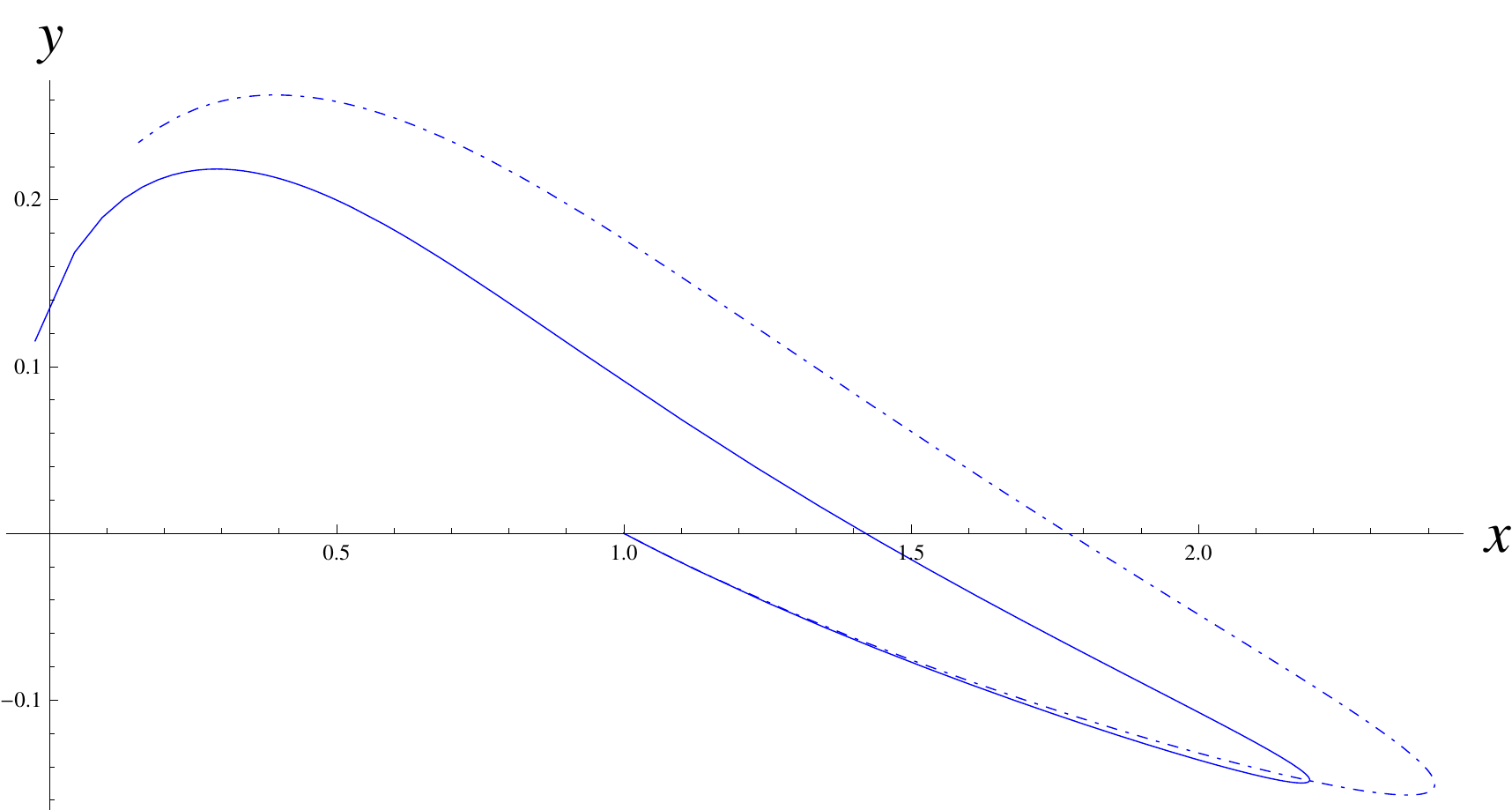}}\\
\caption[]{(a) Effective quantum evolution of $\theta(t)$ and (b)
the quantum modified orbit, for $l=0$ and $m=k=1$, $\Delta l^2 = 0$
solid (blue) curve and $\Delta l^2 \neq 0$ dot-dashed
(blue).}\label{theta-orb-l0}
\end{figure}

However, it turns out that the quantum corrected behavior, up to
second order in the moments, is not reduced to a one effective
dimension when $l=0$. As we can see from equations
(\ref{rdot})-(\ref{gdot}), by setting $l = 0$, and even $\Delta l^n
= 0$, there are still quantum fluctuations related to the angular
momentum as $G^{r, p_{\theta}}$ and $G^{p_r, p_{\theta}}$, from
equations (\ref{g1001}) and (\ref{0101}), which are not necessarily
zero. This is not surprising since, from (\ref{1k}), we see that the
term $G^{r, p_{\theta}}$ is
\begin{equation} \label{1001-k}
G^{r, p_{\theta}}=G^{1,0,0,1}= \left\langle (\hat{r}-r)
(\hat{p}_{\theta}-p_{\theta}) \right\rangle_{\textrm{\tiny{Weyl}}} =
\left\langle \hat{r} \ \hat{p}_{\theta} \right\rangle - lr,
\end{equation}
and similarly for $G^{p_r, p_{\theta}}$; these terms do not vanish.
Indeed, from equation (\ref{theta}) we notice that $\dot{\theta}$
does not vanish but is proportional to $G^{r, p_{\theta}}$. This
implies that quantum fluctuations induce a two-dimensional motion in
the system, modifying the classical behavior of $r$ and
$p_r$, as shown in Fig. \ref{r-p-l0} and \ref{plot1}. The two
dimensional quantum-corrected behavior of $\theta(t)$ is shown in
Fig. \ref{theta-orb-l0}a. In Fig. \ref{theta-orb-l0}b it is clear
that the actual orbit is not one-dimensional but two-dimensional.
This is a purely quantum effect that has no classical parallel.

If we consider $l\neq 0$ in the quantum-corrected equations of
motion, we notice that the system is effectively two dimensional in
the classical sense. If we look at equation (\ref{theta}) we see
that in addition to the term $G^{r, p_{\theta}}$, there are also two other
contributions proportional to $G^{r,r}r^{-4}$ and $r^{-2}$ not present
when $l = 0$. The expectation value of products of linear and
angular momentum operators in general do not commute, therefore the
angular equation does not vanish. We see here that, at the
semiclassical effective level, the classical observables acquire
quantum corrections that are not negligible and that  back-react
latter on the classical variables.

This system does not correspond exactly to the effective description
of the one dimensional Hydrogen atom. To be so all moments related
to angular variables would necessarily equate to zero. Amusingly,
when we impose this in our analysis we notice that the corresponding
solution has qualitatively the same behavior as in Fig.
\ref{r-p-l0}, although in this case the motion is purely one
dimensional.

\subsection{Full two dimensional case}

In standard quantum mechanics energy levels are obtained by solving
the radial Schr\"odinger equation and imposing asymptotic conditions
on the solution \cite{Dirac}. Observables $Q$ are then obtained from
the wave function $\psi(\mathbf{r})$ (for stationary solutions) in
the usual way as its expectation values $Q=\langle \hat{Q} \rangle=
\int \psi^*~ \hat{Q}  \psi \ d^3x$.

The approach based in effective equations treats the dispersions
as quantum corrections to the evolution. Because dispersions also
evolve in time, we must analyze whether our considerations are
valid, i.e., whether the quantum variables can be considered as
perturbations, and under what circumstances. Our main goal is to
study the evolution of expectation values and its deviation from the
classical picture, then we now proceed to solve numerically
equations (\ref{rdot})-(\ref{gdot}).

\begin{figure}[h!]
\centering \subfigure[]{\includegraphics[width=6cm]{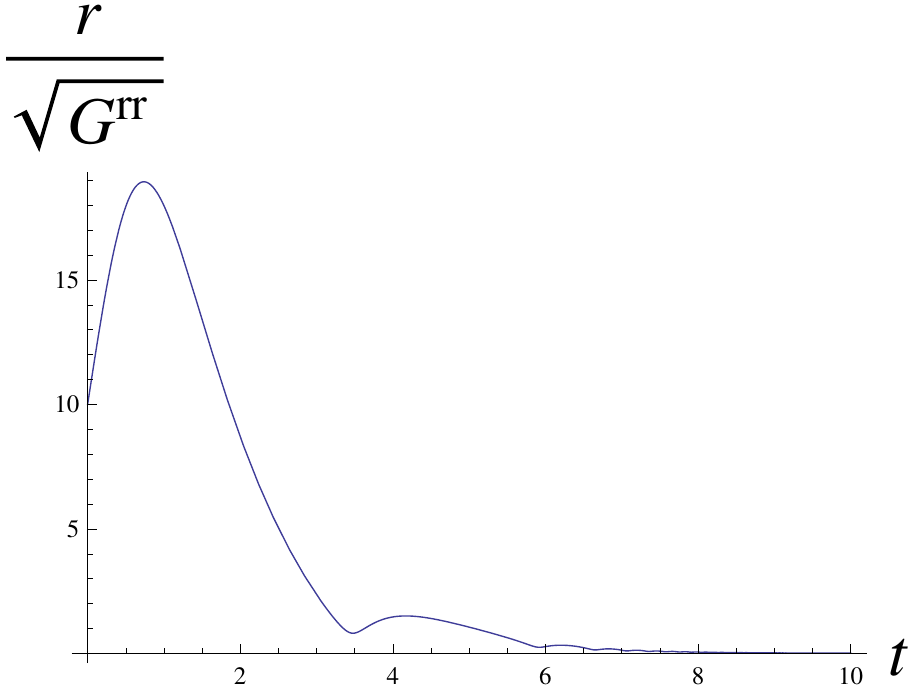}}\hfill
\subfigure[]{\includegraphics[width=6.8cm]{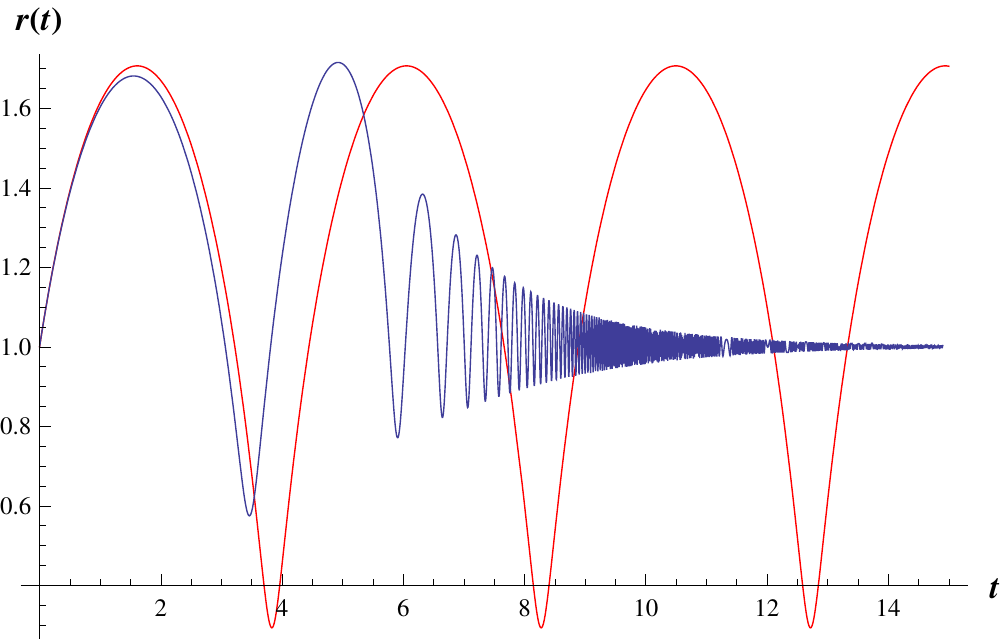}}\\
\caption[]{(a) Plot of $r/\sqrt{G^{rr}}$ with initial conditions
that consider small enough values for dispersions. (b) The radial
coordinate as a function of time. Compared with the classical
behavior, the effective one has a decreasing amplitude as evolves,
and for large times oscillates around $r=1$. In this plots we
consider $m=1$, $k=2$ and $l=1$. The initial conditions for
unperturbed curves are $r=1$, $p_r=1$ and $\theta=\pi$; and for the
moments we consider all of order 0.01.} \label{plot3}
\end{figure}

In Fig. \ref{plot3}a it can be seen that for initial conditions that
consider small enough dispersions, there is a region close to $t=0$
where the dispersions are small compared to the classical variables,
allowing a perturbative evolution. Furthermore, we notice in Fig.
\ref{plot3}b, that for large times the radial coordinate oscillates
with small amplitude around $r=1$. The angular variable and its
dispersion have a similar behavior.

Note that, as the fluctuations evolve, they begin to grow until they
are comparable or even bigger to the expectation
values. In this region they could no longer be considered as
perturbations.

\begin{figure}[h!]
\centering \subfigure[]{\includegraphics[width=6cm]{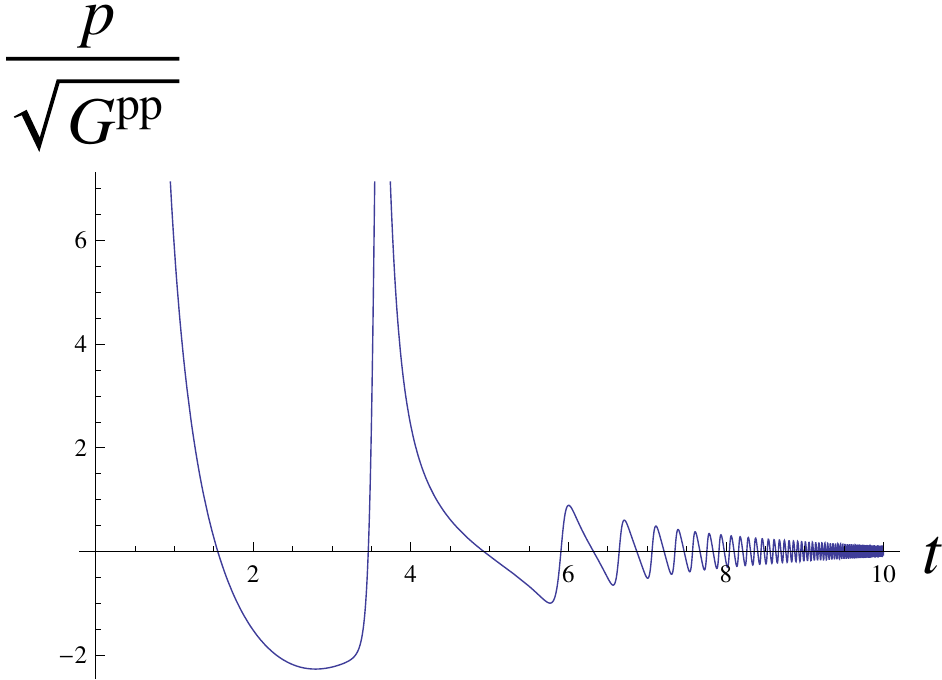}}\hfill
\subfigure[]{\includegraphics[width=6.8cm]{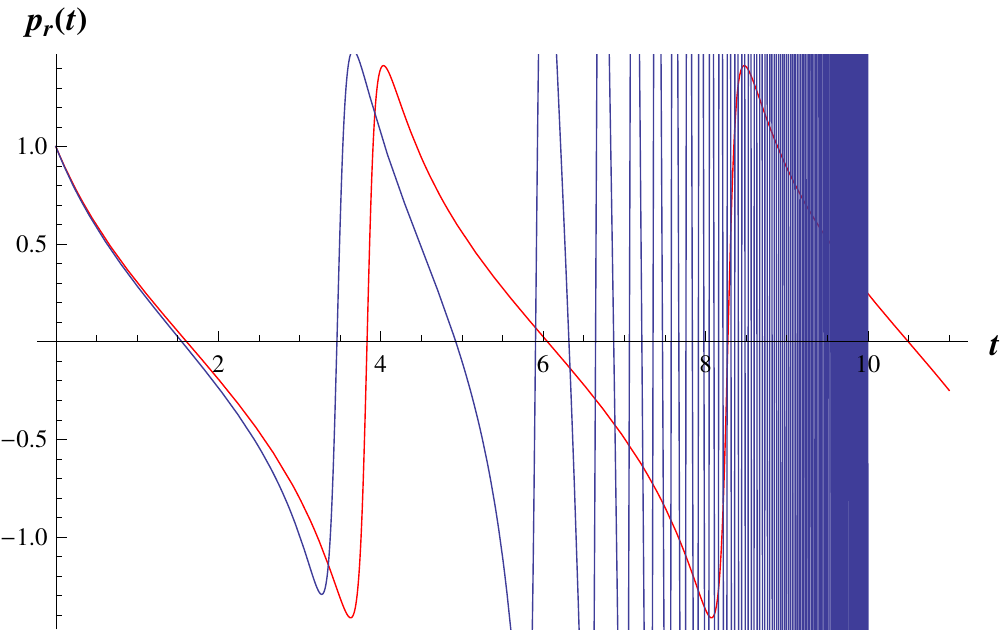}}\\
\caption[]{(a) Plot of $p_r/\sqrt{G^{pp}}$ and (b) of the radial
momentum, both with the same initial conditions as in Fig. \ref{plot3}.}
\label{plot5}
\end{figure}

For  $p_r/\sqrt{G^{p_r p_r}}$ we can see a similar oscillatory
evolution in Fig.~\ref{plot5}, for the same initial conditions. In
the first part  a perturbative approximation can be implemented while for
larger times, it oscillates around zero with small amplitude. The
effective radial momentum as compared with the classical one, starts
close to it but as time increases the momentum becomes larger as
well as its oscillation frequency. Again this implies that for large
times the dispersion are of the same magnitude as the momentum.

\begin{figure}[h!]
\centering \subfigure[]{\includegraphics[width=5.2cm]{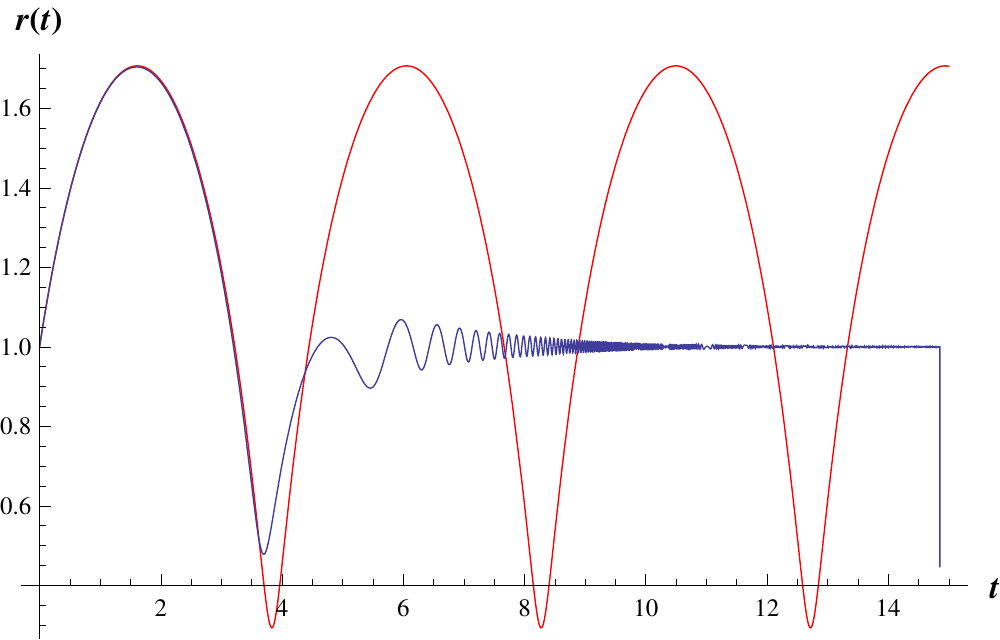}}\hfill
\subfigure[]{\includegraphics[width=5.2cm]{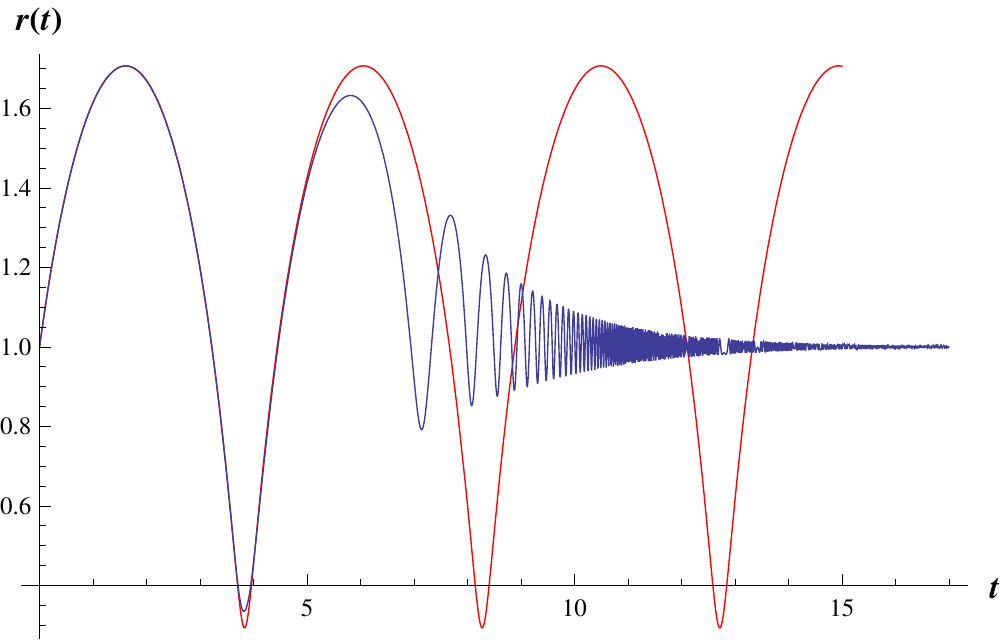}}\hfill
\subfigure[]{\includegraphics[width=5.2cm]{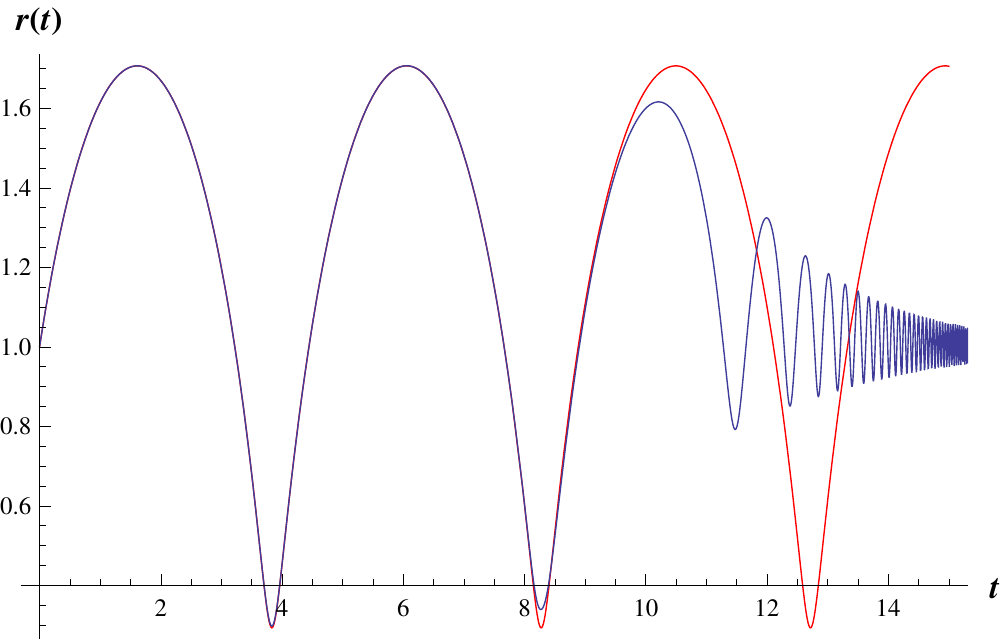}}\\
\caption[]{Plots of the radius as a function of time for initial
conditions fulfilling the uncertainty relation (\ref{2}). (a)
Dispersions of order 0.001, (b) order $10^{-4}$ and (c)
$10^{-5}$. We can see that the smaller the perturbations, the longer
they remain near the classical trajectory, allowing
longer perturbation regions.} \label{plotR}
\end{figure}

\begin{figure}[h!]
\centering \subfigure[]{\includegraphics[width=5cm]{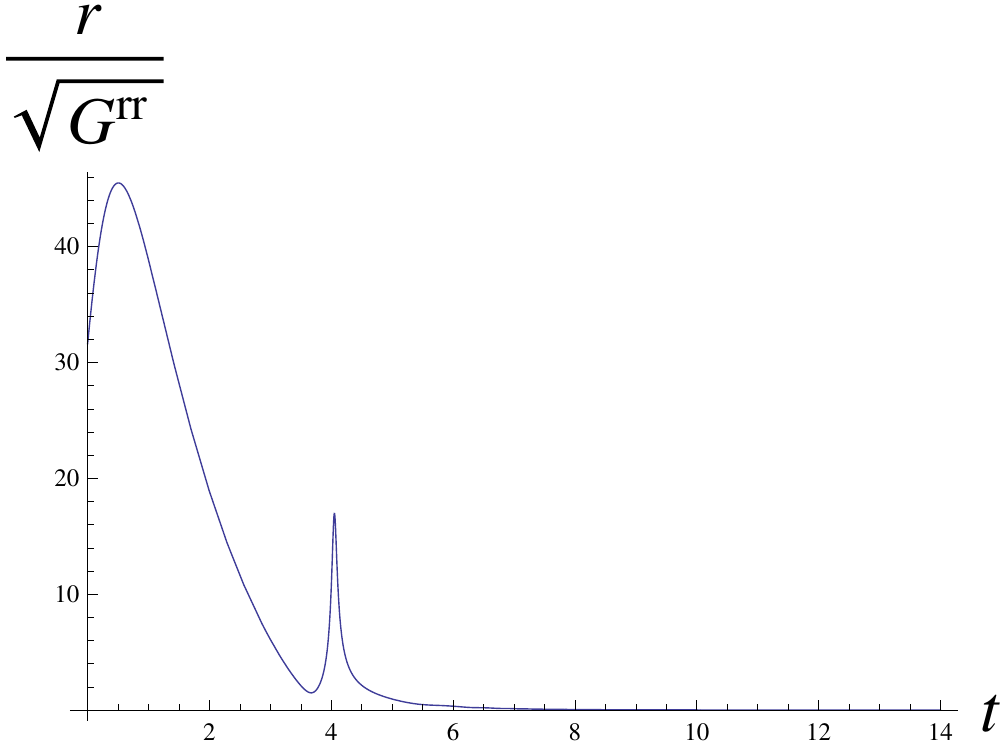}}\hfill
\subfigure[]{\includegraphics[width=5cm]{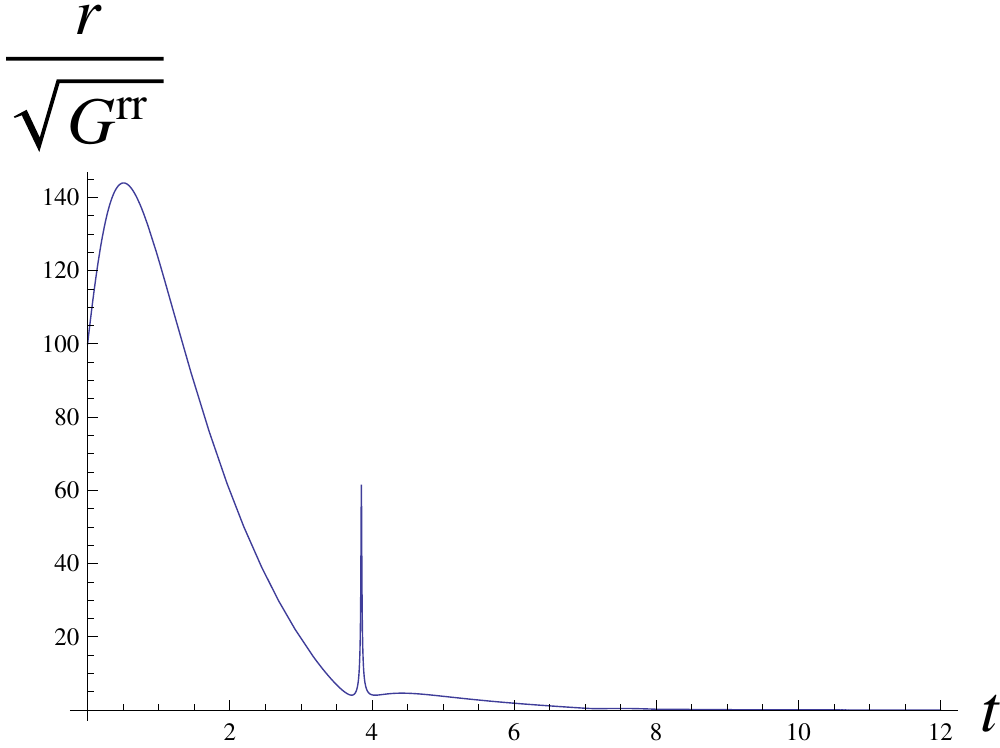}}\hfill
\subfigure[]{\includegraphics[width=5cm]{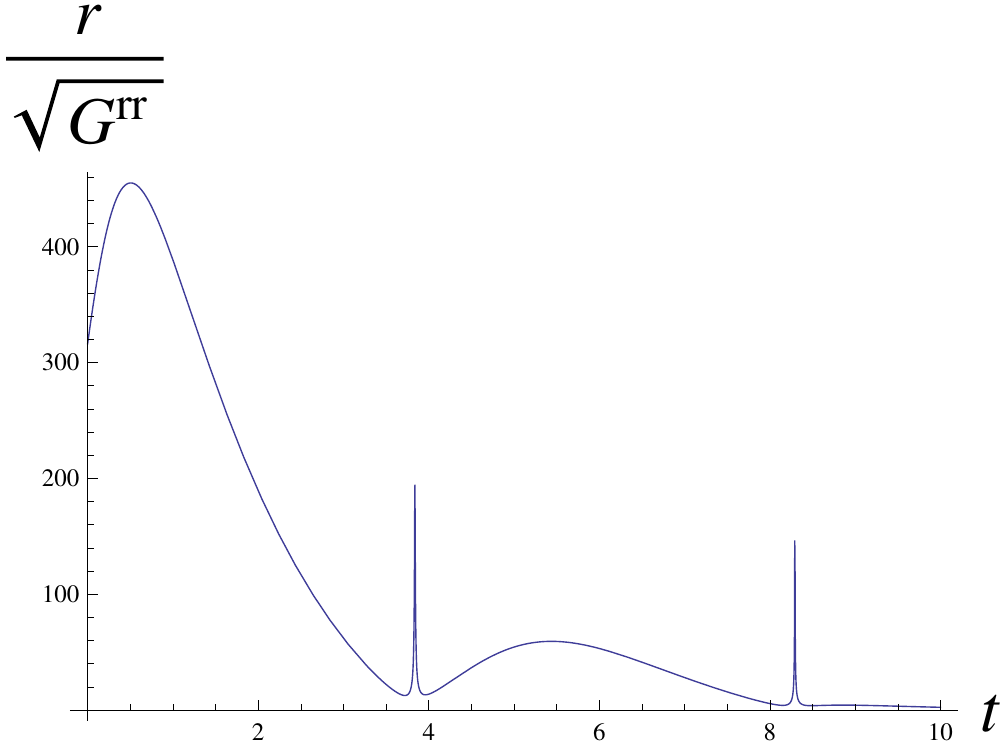}}\\
\caption[]{Plots of the ratio $r/\sqrt{G^{rr}}$ as a function of
time for the same initial conditions as in Fig. \ref{plotR}.
Interestingly, for these initial conditions the graph presents some
peaks which indicates that dispersions may be considered as
perturbations around those regions.} \label{plotRG}
\end{figure}

\begin{figure}[h!]
\centering \subfigure[]{\includegraphics[width=6.4cm]{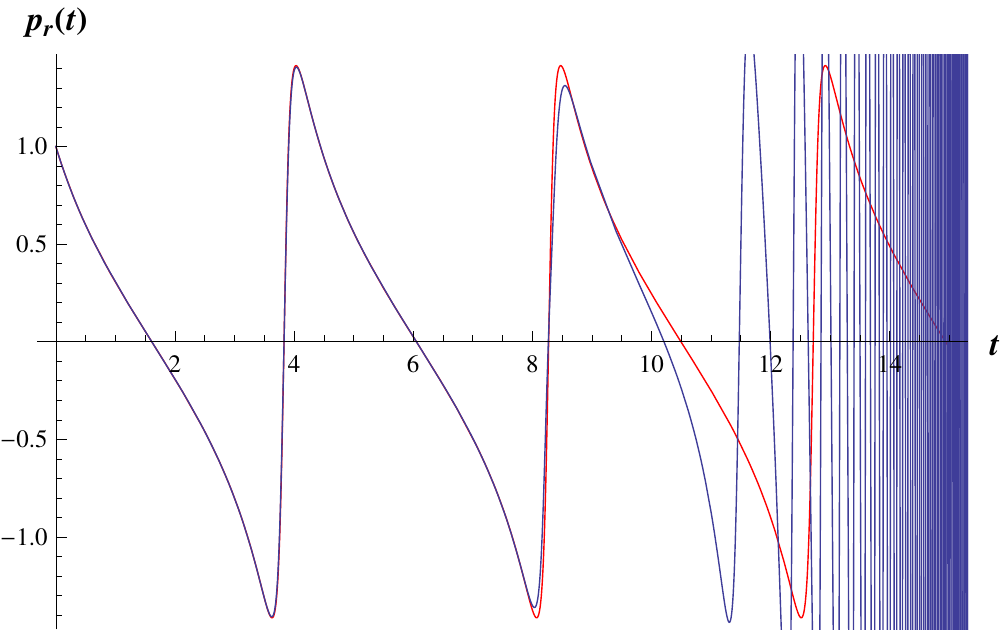}}\hfill
\subfigure[]{\includegraphics[width=6cm]{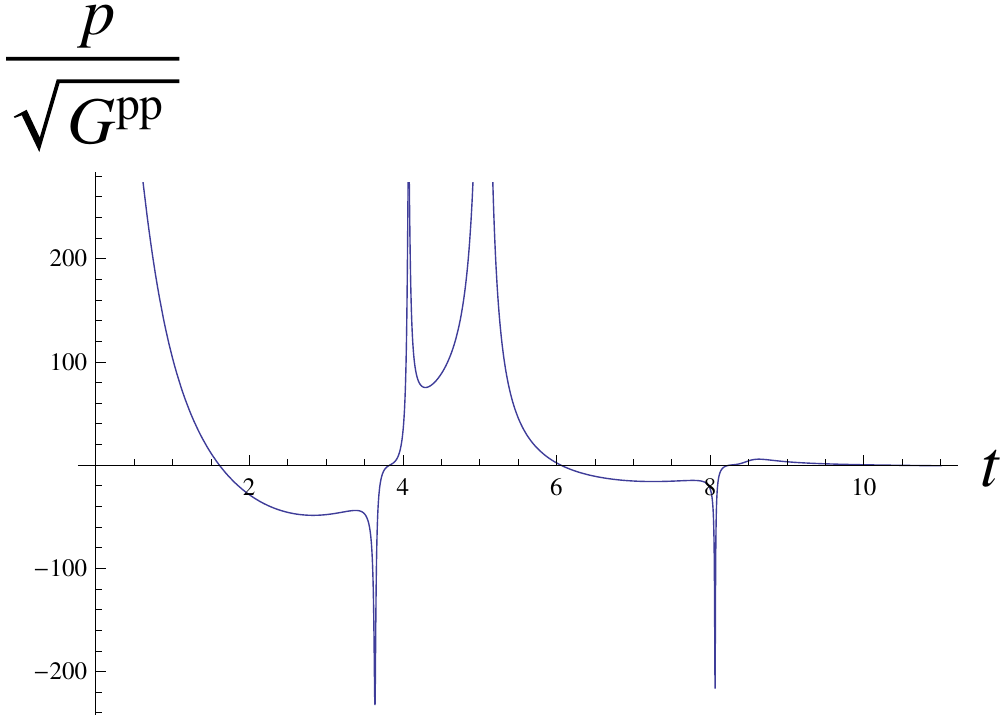}}\\
\caption[]{(a) Plot of $p_r$ and (b) $p_r/\sqrt{G^{pp}}$, for
initial conditions of dispersions of order $10^{-5}$. There are
piecewise regions where the perturbative approximation is well suited.}
\label{plotPG}
\end{figure}

One can always impose initial conditions
 complying with the uncertainty principle (\ref{2}) for each
pair of conjugate variables. We can see this  in Fig.~\ref{plotR}
for the radius, Fig.~\ref{plotRG} for the ratio of the corresponding
dispersions and Fig.~\ref{plotPG} for their momenta, taking for the
dispersions initial conditions progressively smaller. Although the
behavior is qualitatively similar to that shown in Fig.~\ref{plot3}
and Fig.~\ref{plot5}, we can notice that, for the initial conditions
that fulfill the uncertainty relations, we have better control of
the approximation. This is because the effective evolution is very
close to the classical one. Indeed, the smaller the
initial condition for the perturbation, the longer the effective
trajectory remains close to the classical one. One can see
this kind of behavior, where the classical orbits decay after some
time, when one considers the hydrogen atom interacting with
external fields \cite{cole}.

%

\begin{figure}[h!]
\centering
\includegraphics[width=6.8cm]{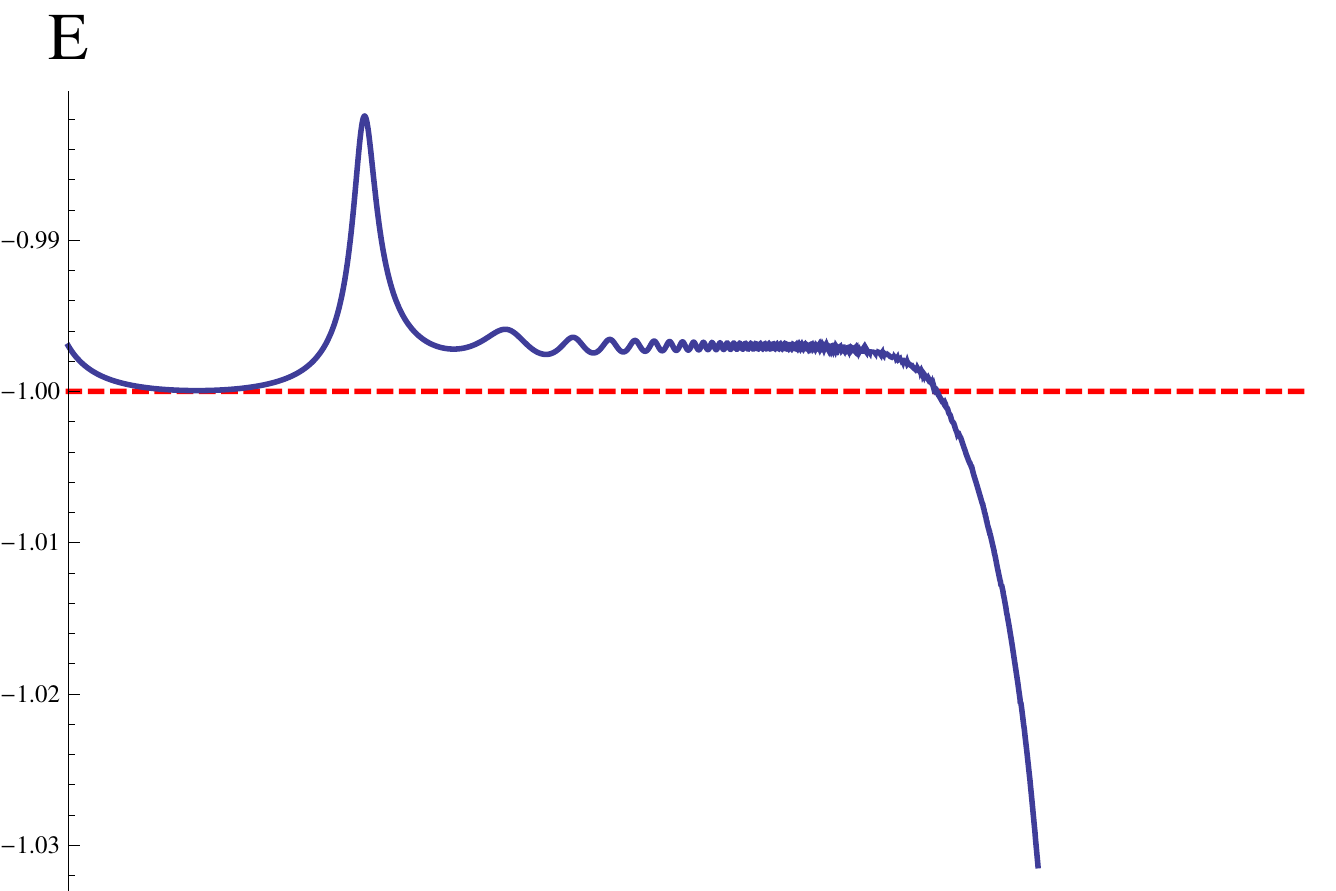}
\caption{The classical (dashed red line) and quantum corrected
(solid blue line) energy as function of time. The initial
conditions for dispersions are of order 0.001.} \label{plot6}
\end{figure}

It is expected that the energy of the system is also back-reacted,
which indeed is verified from the definition of the effective
quantum energy (\ref{10}). However, as classically the energy is a
constant of motion, in the perturbative regime the effective
behavior of the corrected energy oscillates about the classical
energy with small amplitudes that depend on the initial conditions
for the dispersions, as shown in Fig.~\ref{plot6}.
%

\begin{figure}[h!]
\centering \subfigure[]{\includegraphics[width=6.2cm]{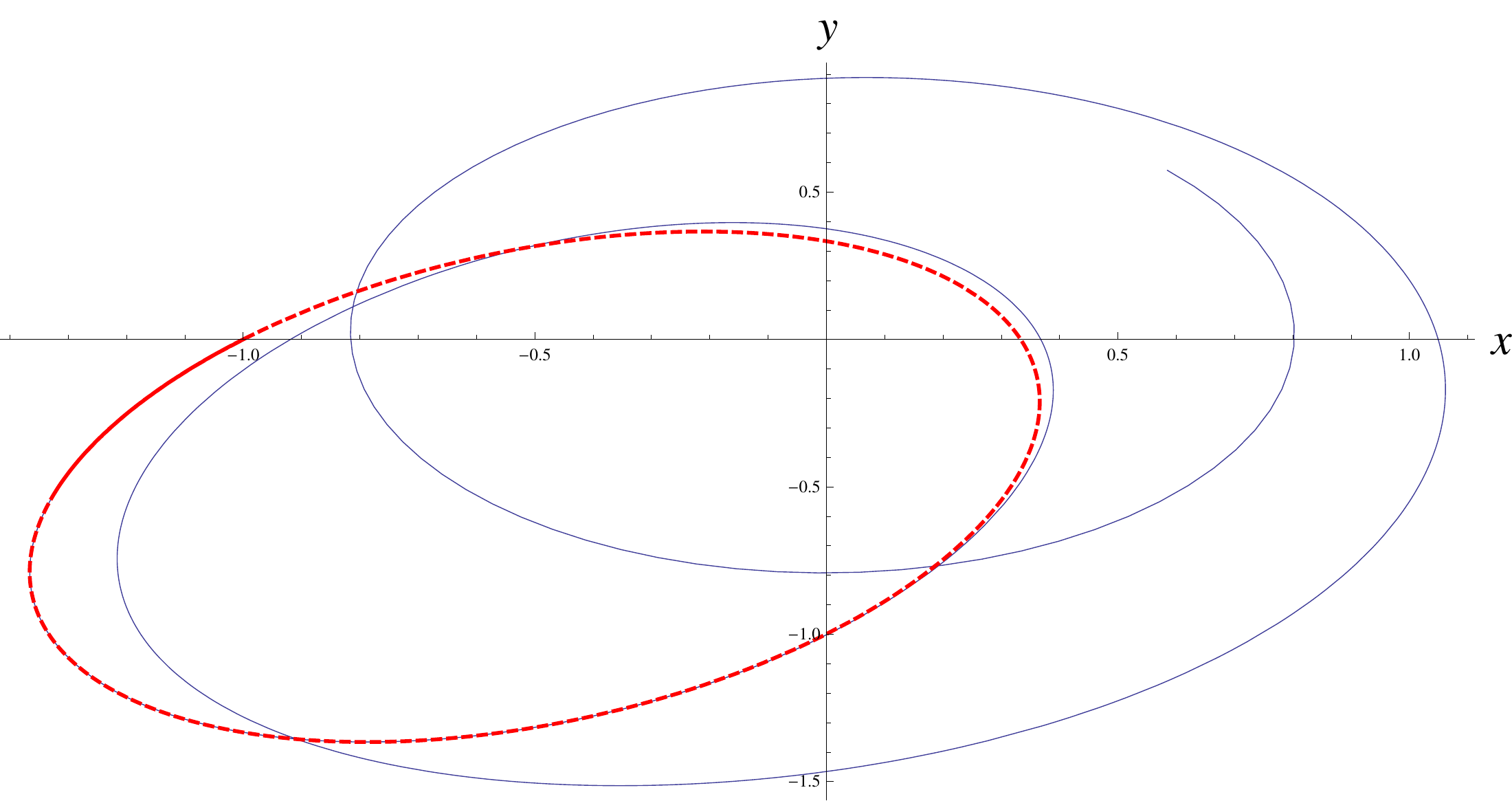}}\hfill
\subfigure[]{\includegraphics[width=6.2cm]{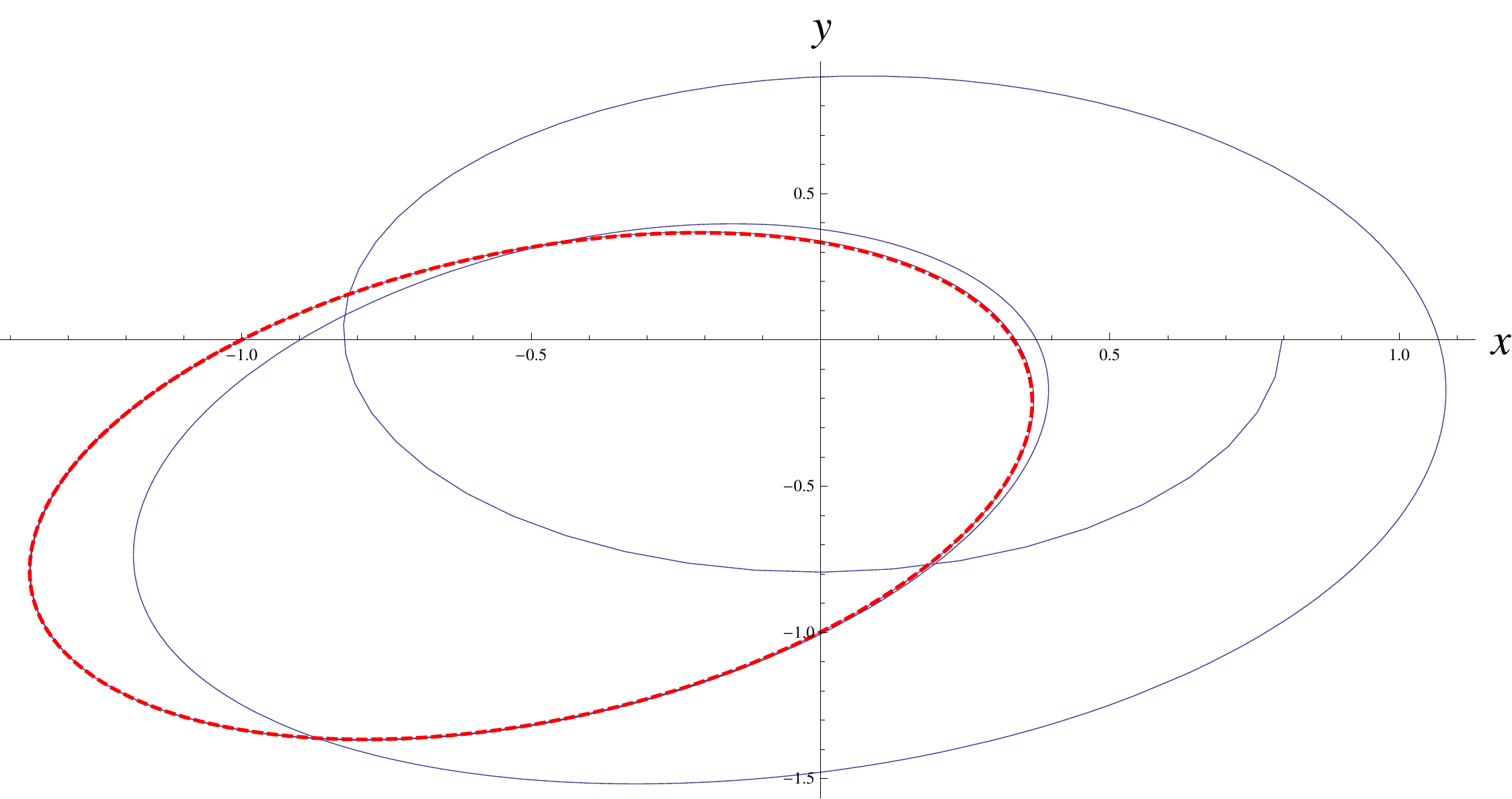}}\\
\caption{The classical (dashed red line) and quantum corrected
(solid blue line) orbits for dispersions of order (a) $10^{-4}$
and (c) $10^{-5}$. Classical and quantum-corrected orbits starts close,
 but at large times the effective one spreads away.} \label{orbit}
\end{figure}

Let us recall that classically the solution of the Kepler problem
corresponds to a conic-section orbit whose eccentricity and
perihelion depends on the energy and angular momentum. Although in
the quantum-corrected study that we are discussing (as in any
quantum/statistical description) one cannot have a closed analytical
expression for the effective orbit, one can solve the system
(\ref{rdot})-(\ref{gdot}) numerically. With initial conditions
fulfilling the uncertainty relations, we obtain an almost elliptical
(open) orbit that is modified by the effect of back-reaction of
quantum variables, as shown in Fig.~\ref{orbit}. Under these
conditions the effective orbit starts very close to the classical
one and, as it evolves, quantum effects increasingly drive it away.

\begin{figure}[h!]
\centering
\subfigure[]{\includegraphics[width=5.3cm, height=4.2cm]{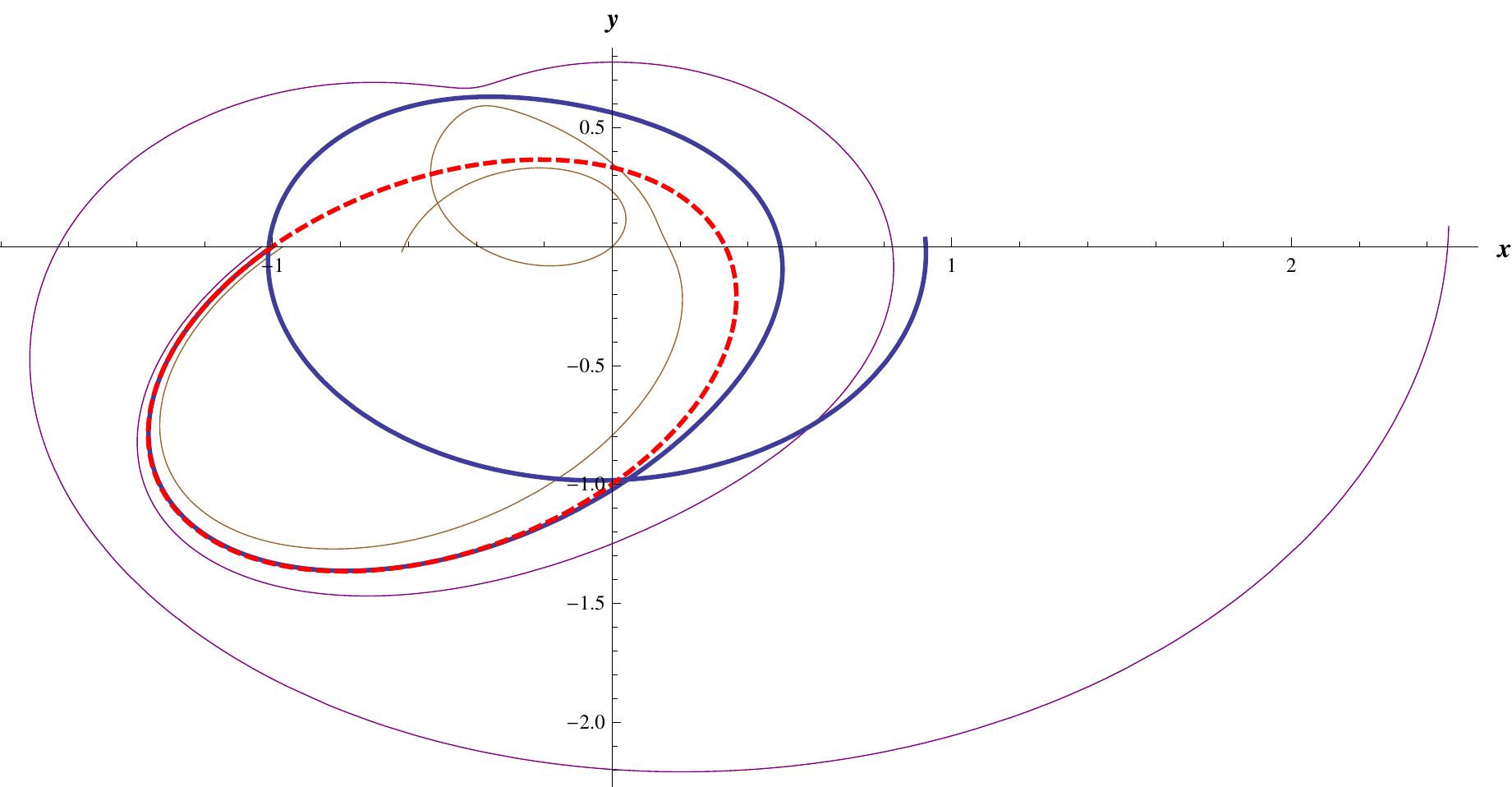}}\hfill
\subfigure[]{\includegraphics[width=5.2cm, height=4.1cm]{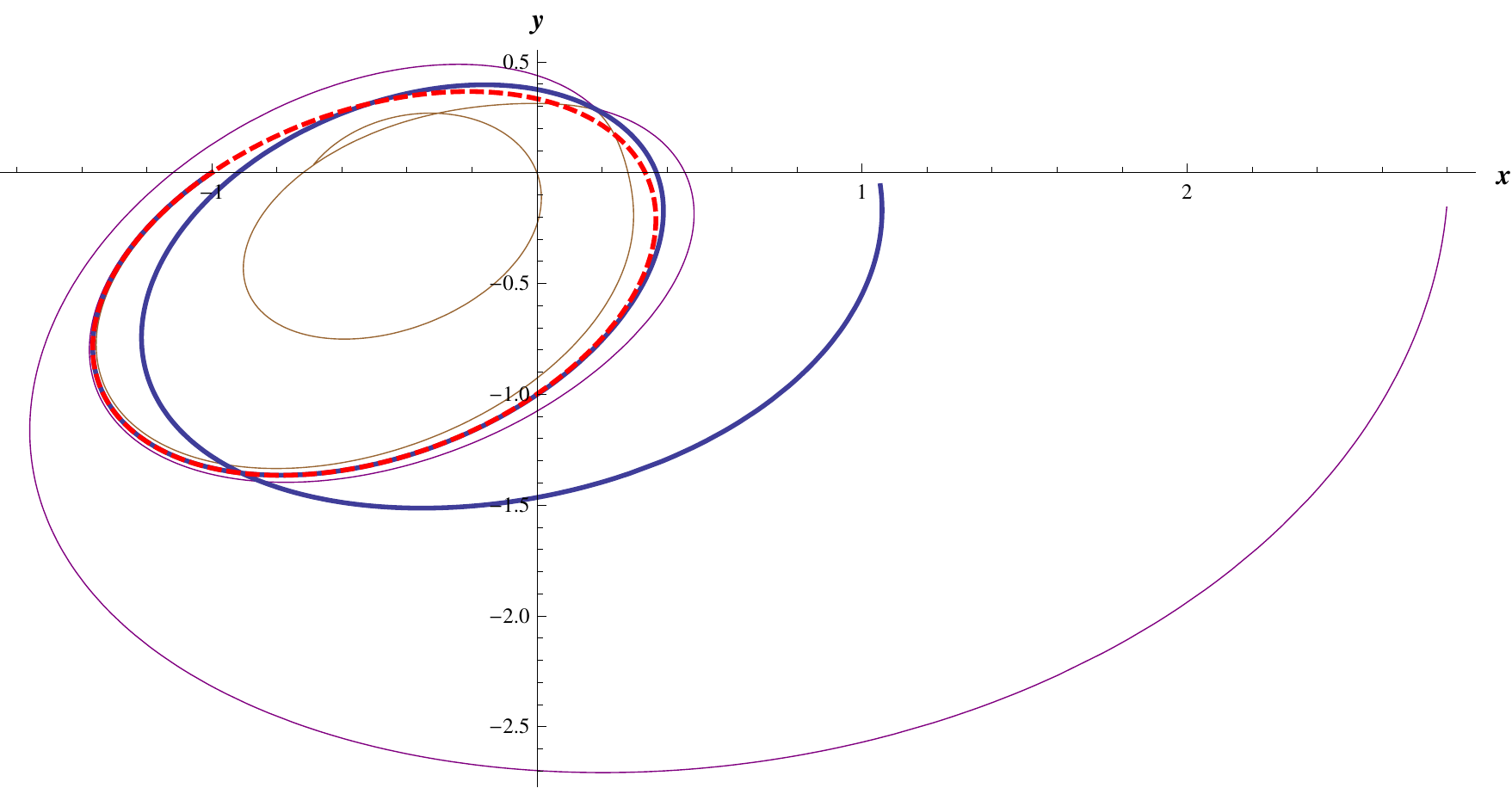}}\hfill
\subfigure[]{\includegraphics[width=3.8cm, height=4.3cm]{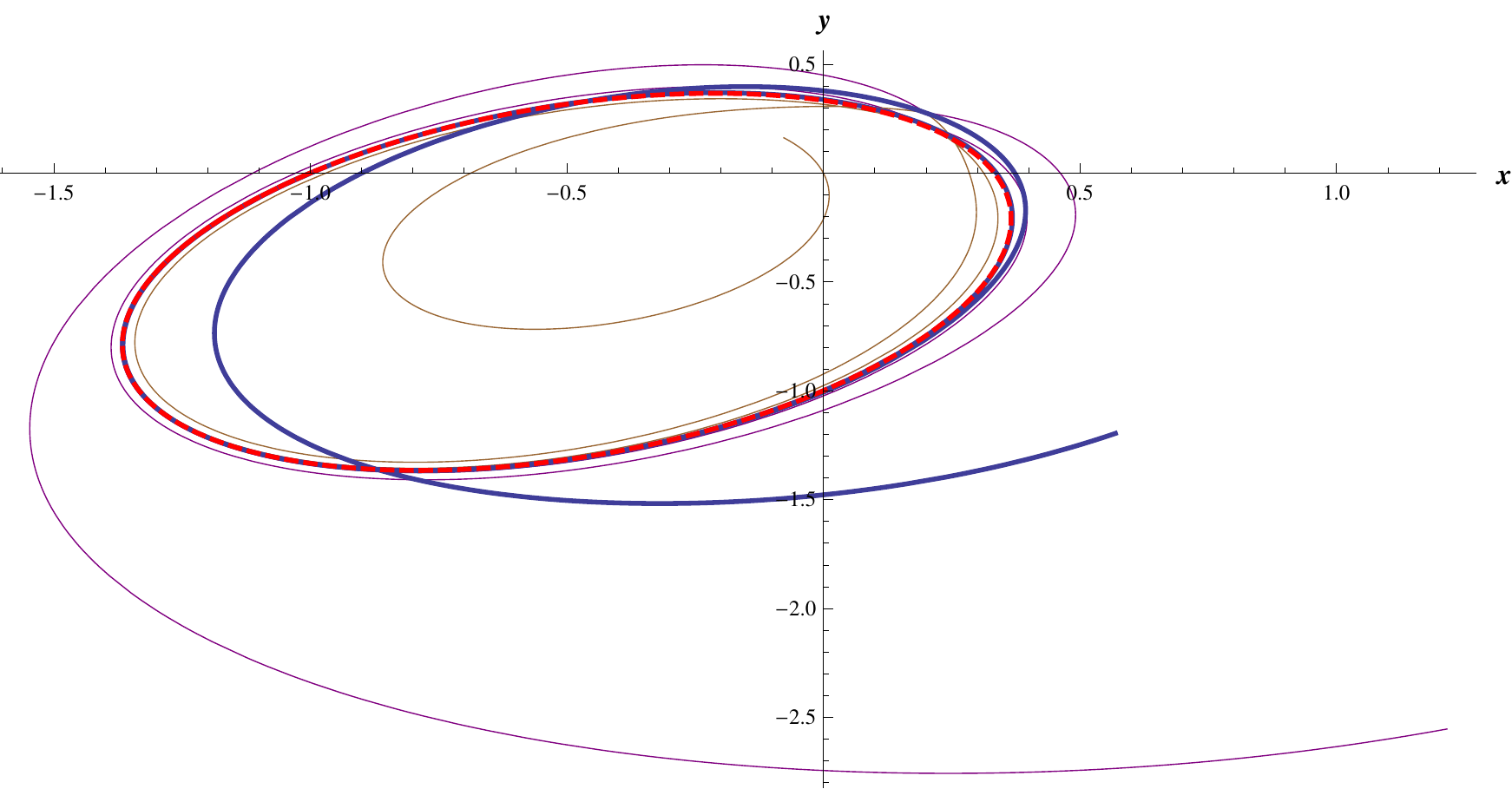}}\\
\caption[]{Plot of the quantum-corrected (blue) orbit surrounded by the
corresponding dynamical uncertainties $r \pm \Delta r$ (thin
curves), for initial conditions of the dispersions of order (a) 0.001,
(b) $10^{-4}$ and (c) $10^{-5}$. Also the classical orbit (red
dashed curve) is shown. Initially the uncertainties are close
to, and at each side of, the effective orbit, then they start to disperse.
 At large times they diverge completely from the
classical behavior.} \label{orbitD}
\end{figure}

The departure of the effective orbit from the classical one can also
be understood by looking at the uncertainties in $r$.
These are the curves located at each side of the quantum trajectory $r(\theta)$, (i.e. the brown and purple thin curves).
in Fig.~\ref{orbitD}. Here we notice how,
at the beginning of evolution, both the effective and classical
orbits are within the range of dispersions $\pm\sqrt{G^{rr}}=\pm
\Delta r$. As the system evolves we can see how the uncertainties
start to grow and disperse. Interestingly at some point
the uncertainties decrease and we recover the effective trajectory.
For large initial conditions of the uncertainties this does not happen very close to the classical
trajectory, but as one decreases the initial values this happens
 closer to the classical behavior,  Fig.~\ref{orbitD}. Actually,
from the previous discussion we know that there is a region where
the magnitude of the dispersion is comparable to that of the
expectation values. Then we can interpret these as the quantum
effects that keep the orbit away from the classical trajectory for
large times. This behavior is similar to the one in
\cite{stroud}, where minimum uncertainty states were constructed in
which the wave function can be described by classical equations for
short times, while for longer times the wave packet that lives in
the elliptical trajectory begins to spread.

\begin{figure}[h!]
\centering \subfigure[]{\includegraphics[width=7cm]{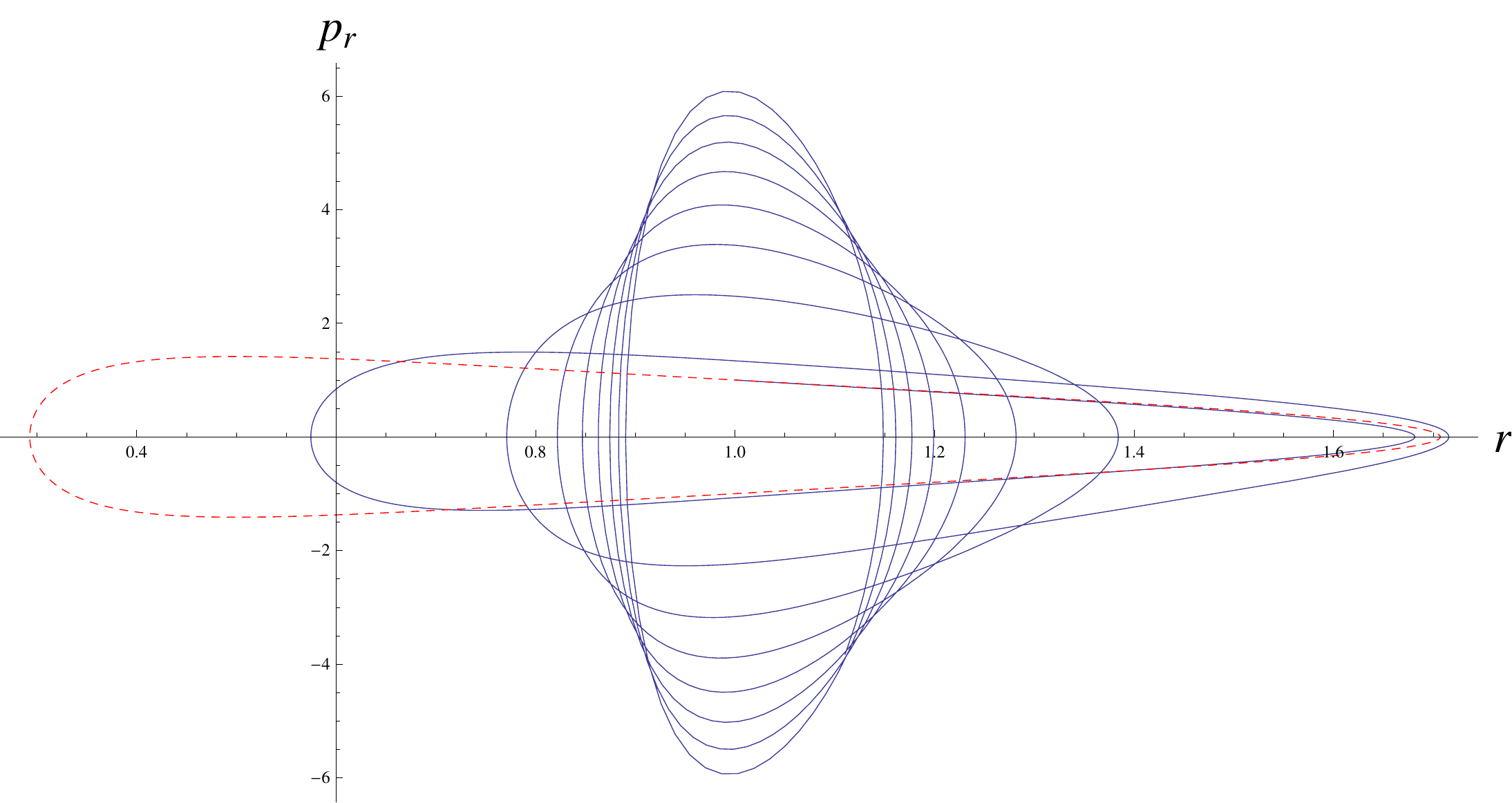}}\hfill
\subfigure[]{\includegraphics[width=6.6cm]{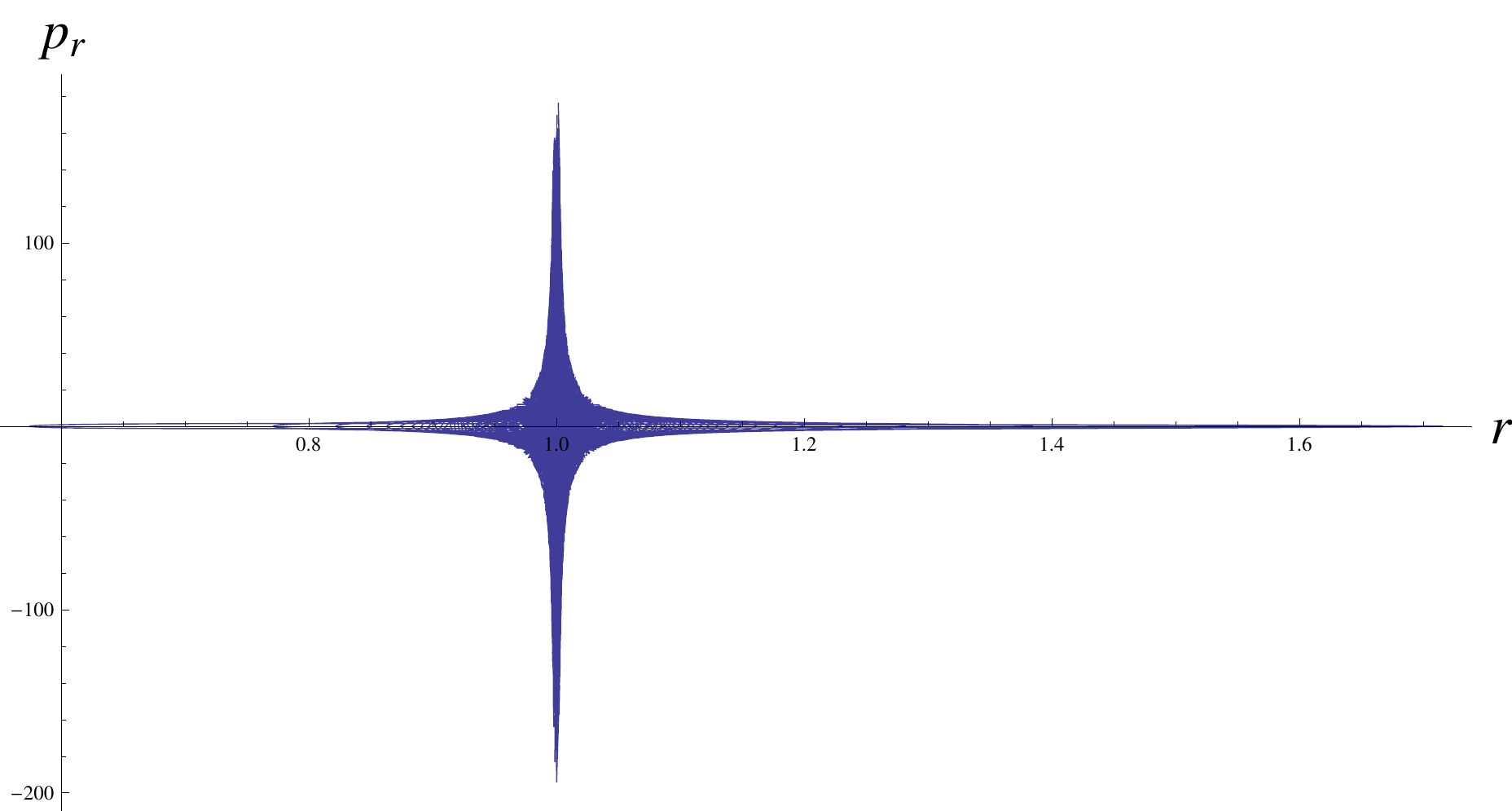}}\\
\caption[]{(a) The dashed (red) line corresponds to the classical
radial phase diagram while the solid (blue) curve is the
effective quantum diagram for initial conditions of dispersions of
order  0.01. (b) For later times the radial coordinate
is fully localized while the radial momentum disperse completely.}
\label{plot4}
\end{figure}

We can further appreciate the quantum evolution of the system by
looking at the corresponding radial phase space diagram.
Classically, for a particle rotating in an elliptic orbit in
configuration space, the radial phase diagram is a closed curve
representing a bound state in which the particle attains a maximum
and a minimum value of its radius from the force center (dashed (red) curve
in Fig.~\ref{plot4}a). The solid curve in
Fig.~\ref{plot4}a corresponds to the quantum-corrected evolution; we
notice that the effective quantum behavior starts close to the
classical diagram and then disperses. This diagram shows how the
quantum effects make the effective system depart from the classical
one. As time evolves we see that the radius becomes increasingly
smaller and the radial momentum grows. Moreover, after a long enough
period of time, we can see in Fig.~\ref{plot4}b that, as the radius gets
localized, the radial momentum disperses: this is a direct
manifestation of the uncertainty principle in phase space.

\begin{figure}[h!]
\centering \subfigure[]{\includegraphics[width=6cm]{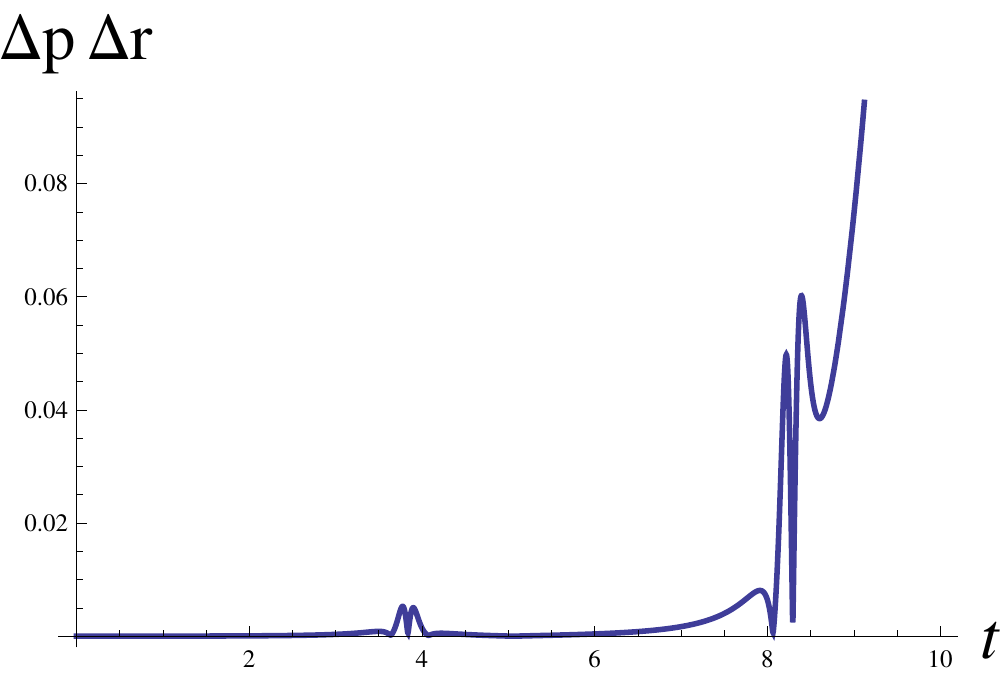}}\hfill
\subfigure[]{\includegraphics[width=6.2cm]{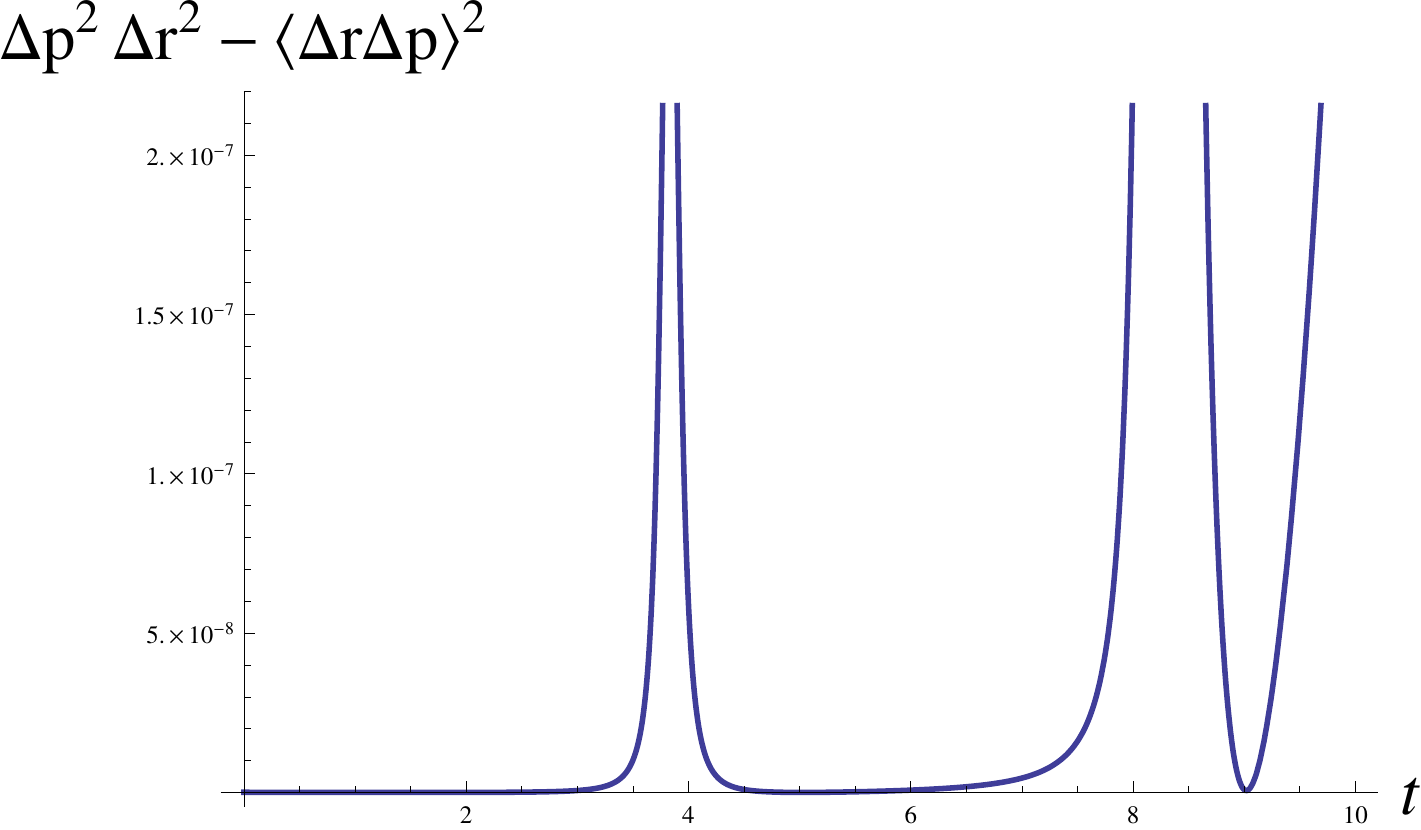}}\\
\caption{The uncertainty relations (a) $\Delta r\Delta p_r$ and (b)
$G^{rr}~ G^{p_r p_r} - (G^{rp_r})^2$ as function of time (order
$10^{-5}$). During the perturbative stage, the uncertainty relations
are satisfied; for larger times the inequality is violated
 and quantum effects become increasingly dominant.}
\label{plot8}
\end{figure}

Fig.~\ref{plot8} shows how the dispersions evolve and to what extent
the uncertainty principle is fulfilled. From Fig.~\ref{plot8} we see
that the inequality is satisfied for short times, during
the phase where the expansion can be treated
perturbatively, as mentioned above. There is only one region where
the product presents a peak but then remains flat for some more
time. For large times the perturbative assumption is no longer valid, as the
product of uncertainties grow larger and larger. This is a similar behavior as
the one presented by a radial squeezed state studied in
\cite{kosteleky}, where the uncertainty product features a cyclic
behavior. In the same way we can say that our system begins in a
state with minimum uncertainty, not squeezed, and for larger
times it gets dispersed.
%


%
\subsection{Different approaches to semi-classical limit of the Hydrogen atom}

Even though the non-relativistic Hydrogen atom has been solved
exactly it was latter proved that the WKB approximation was rather
poor in this case, which is usually attributed to the singular
behavior of the potential at the origin $r = 0$. It was shown in
\cite{wkb} that WKB can be performed in $r \rightarrow 0$ with
precise results if the centrifugal term
\[ V_C= \frac{ \hbar^2 l (l+1)}{2mr^2} \]
is replaced by
\[ V_C= \frac{ \hbar^2  (l+1/2)^2}{2mr^2} \]
which in turn gives very accurate energy eigenvalues for the WKB
approximation, even at lowest order. The main difference between WKB
and the method presented here is the scope of the treatment: while
both methods intend to establish a semiclassical approach to
the quantum mechanical system at hand, the WKB aims at obtaining, by
means of an series expansion on $\hbar$, the wave function and
corrections to the energy levels. That is, one
writes the wave function as \cite{Q}
\[ \psi(\mathbf{r})=e^{\frac{i}{\hbar} W(\mathbf{r})}, \]
and the WKB approximation consists in expanding $W(\mathbf{r})$ in
powers of $\hbar$ and in neglecting in the Schr\"odinger equation
terms of order $\hbar^2$ or higher. Besides the fact that the method
aims in approximating the wave function it fails in the vicinity of
the \emph{classical} turning points, i.e., $E=V(x)$. To circumvent
this one defines the \emph{connection} functions in order to connect
the wave functions in the regions adjacent to the turning points.

Evolution and description of the quantum mechanical system in our
method is not based on a wave function but rather in its momentous
description and their quantum back-reaction on expectation values.
One can compute the energy of the system and, probably the most
interesting aspect of our description, also the semiclassical
evolution of observables and orbits at different regimes in the
parameter space. However, one feature that both schemes (and any
approximation method in quantum mechanics) share is the growing in
time of quantum corrections that render the approach invalid as well
as the oscillatory character of the evolution in our case or in the
expression of wave function and energy in the WKB. This is a well
known effect in quantum mechanics.

The effective equations method presented here is an alternative
method that in principle should reproduce all the results of
standard quantum mechanics as long as one consider all the infinite
moments. As mentioned, this method is quite suitable for a
semi-classical analysis, defining a region where the moments can be
considered as perturbations.

The semi-classical states of the Hydrogen atom have been explored in
terms of the correspondence principle for large principal quantum
number $n$. In such a case a wave packet can follow a classical
orbit. Interestingly, given the conservation of energy and angular
momentum, there are no transversal spreading in such states. The
simplest of this states has a dispersion of order $n^{-1/2}$, and it
evolves in such a way that the wave packet is distributed almost uniformly on a
circular orbit \cite{brown}. The dynamics of these states have both
classical and quantum features: for short times the motion is
classical, after that the quantum dynamics becomes dominant
\cite{stroud}. Something very similar happens in our case, although
we do not necessarily describe the same state, as the effective
equations method provides us with minimal uncertainty dynamical
states. There is also a formalism to construct radial and angular coherent and squeezed
states that show the main features of
classical motion \cite{kosteleky, nieto}. Semi-classical states have
also been constructed from the coherent states of a 4 dimensional
oscillator which is reached by the KS mapping of the 3 dimensional
Hydrogen atom \cite{Ghosh}.

It is interesting to study the semi-classical states of the hydrogen
atom as it has been possible to find them experimentally in the
laboratory through the excitation of the outermost electrons of
certain atoms, with ultrashort laser pulses \cite{exps1}. Recently,
there have been obtained experimentally new localized wave packets
in very large $n$ states that travel in nearly circular orbits
\cite{exps2}.

\begin{figure}[h!]
\centering
\includegraphics[width=6.6cm]{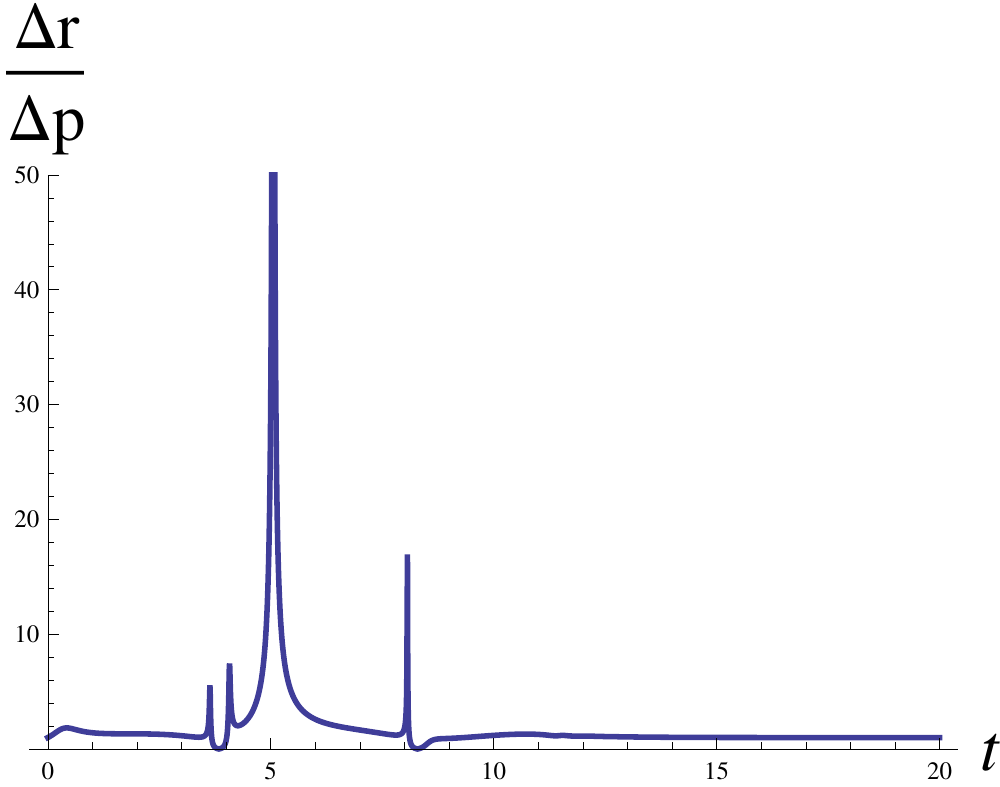}
\caption{Ratio $\Delta r / \Delta p_r$. The peaks indicate that
there is more spread in the radius than in the momentum at that
time, i.e. states are squeezed for short periods.}\label{Dratio}
\end{figure}

The analysis presented here provides an alternative tool for the
study of some different semi-classical states of the Hydrogen atom
that had not been considered before, indeed we can see that our
states are not of minimum uncertainty for all times. This is even
clearer if we consider the ratio $\Delta r / \Delta p_r$ in Fig.
\ref{Dratio}. One can see that there is a group of peaks indicating
that at those times the uncertainty in $r$ increases as
uncertainty in $p_r$ decreases, while in the flat regions the
opposite occurs. As in Fig. \ref{plot8} we can compare this with
the behavior of the radial squeezed state \cite{kosteleky}, where
this ratio starts with a large amplitude and then damped oscillates
until it reaches a minimum although continues oscillating, showing the
squeezing in $p_r$. Clearly our state is not a squeezed one, but is
the closest one that begins satisfying uncertainty relations.

\section{Discussion}\label{Discussion}

In this work we have obtained an effective description of the
Hydrogen atom by means of a momentous formulation of quantum
mechanics. We were able to obtain a semiclassical picture of the
evolution of the classical dynamical variables by studying their
equations of motion, back reacted by quantum dispersions and
spreadings. This is done in regions where the moments can be
considered as perturbations, which also allows us to make a
consistent truncation of the equations of motion.

The equations of motion for expectation values are obtained from an
effective Hamiltonian that acquires quantum corrections represented
by spreadings to all orders. We have seen that the system reduces to
the Kepler problem when one sets all the quantum fluctuations to
zero. The case of zero angular orbital momentum, $l=0$, that
classically corresponds to a one dimensional system, gets so
strongly quantum corrected that it effectively corresponds to a two
dimensional system. This behavior has no classical analogous, and in
this way is a similar effect as the quantum tunneling phenomena.

The evolution of the system is controlled by generalizations of the
Heisenberg uncertainty relations among quantum variables that
are satisfied during the perturbative period,
 thus providing us with a consistent analysis and evolution.

As we pointed out we were able to determine the evolution of
expectation values of variables which enabled us to determine an
effective orbit for the electron, which is close to the
classical elliptical orbit for short initial times. For larger times it is
diverted to an open quasi-elliptical orbit as shown in Figs.
\ref{orbit}, \ref{orbitD} and \ref{plot4}. One can see from these
figures how quantum effects modify the classical behavior,
realizing that, as the system evolves, there is a switching
between the radial and momentum spreadings. It is interesting to point
out that the behavior of the orbits and the spreadings shown in Fig.
\ref{orbitD}, although not identical, are similar to those reported
in the literature where the wave packets follow elliptical orbits
for a while and then spread.

Although this method gives a semi-classical behavior in a
well-defined region, the states that we use are not the minimum
uncertainty states that are usually called coherent states for the
Hydrogen atom, or the Rydberg-type states similar to those that have
been experimentally produced through ultra-short laser pulses. Our
states are new states describing the semiclassical behavior of the
Hydrogen atom. Thus we emphasize the qualitative features of this
analysis, the method of effective equations and its robustness in
obtaining new states suited to describe the semiclassical behavior.

Even though we were able to solve
numerically the equations of motion, and with this
obtaining the evolution of the
system providing a rather interesting description, it is still
needed to obtain a set of
 physical initial conditions that will allow us
to compare quantitatively our results with experimental or
theoretical data in a consistent manner. Due to the very complicated nature of the
dynamical system and the infinite number of quantum variables this
task is yet under further study.

Due to the complexity of the effective variables and their equations
of motion we restricted ourselves to the two dimensional case,
corresponding to the classical version, and then it is necessary to
explore the three dimensional system. Given the results obtained in
the one dimensional case, where quantum effects induce a two
dimensional behavior,  we could obtain modifications to the two
dimensional motion due to quantum back-reaction by considering the
remaining angular variable and its associated momenta, fixed in the
present analysis. In such case the corresponding results should be
more accurate and closer to those currently reported in the
literature and experimentally.

The implications and possible applications of our analysis to other
systems and phenomena are promising. For instance, the
implementation of this method to Hydrogen-like atoms and systems should be
straightforward and may result in a more detailed description and a deeper understanding
of those systems. Note that, even for the Helium atom (a system
with two electrons), there is no exact solution, and any
study of this system is performed at an approximation level,
being ours one that could led to a more refined and precise
description.

\section*{ACKNOWLEDGEMENTS}

This work was supported by grant CONACYT CB-2008-01/101774.

\appendix

\section{Higher order moments}

As shown in \cite{D}, as one considers higher order moments the
average number of terms in the equations increases exponentially.
Here we show the system of effective evolution equations up to third
order, that corresponds to thirty coupled equations
\begin{eqnarray}
  \fl \dot{p_r} = \frac{l^2}{mr^3}  - \frac{k}{r^2}   + \frac{\Delta l^2}{mr^3} +  \frac{3}{r^{4}}G^{2,0,0,0}\left[ \frac{l^2}{m}\frac{2}{r} - k
  \right] -\frac{4}{r^{5}}G^{3,0,0,0}\left[ \frac{l^2}{m}\frac{5}{r} - k
  \right] \nonumber \\
   - \frac{6}{mr^{4}}\left(l G^{1,0,0,1} + \frac{1}{2}G^{1,0,0,2}\right) + \frac{12\,l}{mr^5}G^{2,0,0,1},\\
  \fl \dot{\theta} = \frac{l}{mr^2} - \frac{2}{mr^3}G^{1,0,0,1} +
    \frac{3}{mr^{4}}\left(lG^{2,0,0,0} +G^{2,0,0,1} \right) - \frac{4l}{mr^{5}}G^{3,0,0,0}, \\
  \fl \dot{G}^{1,1,0,0} = -\frac{1}{m} G^{0,2,0,0} + \left[\frac{3l^2}{2mr} - k
  \right]\frac{2}{r^{3}}G^{2,0,0,0} - \left[\frac{2l^2}{mr} - k
  \right]\frac{3}{r^{4}}G^{3,0,0,0} \nonumber\\
   - \frac{2}{m r^{3}}\left(lG^{1, 0, 0, 1} + \frac{1}{2}G^{1, 0, 0, 2}\right) + \frac{6 l G^{2, 0, 0, 1}}{m r^4}, \\
  \fl \dot{G}^{0,2,0,0} =  4\left[\frac{3l^2}{2mr} - k
  \right]\frac{1}{r^{3}}G^{1,1,0,0} - 6\left[\frac{2l^2}{mr} - k
  \right]\frac{1}{r^{4}}G^{2,1,0,0} -\frac{2}{m r^3}\left(2 l G^{0, 1, 0, 1} + G^{0, 1, 0,
  2}\right) \nonumber\\
   + \frac{12 l}{m r^4} G^{1, 1, 0, 1}, \\
  \fl \dot{G}^{0,0,1,1} =  -\frac{\Delta l^2}{mr^2}   + \frac{2}{m r^{3}}\left(lG^{1, 0, 0, 1}+G^{1, 0, 0, 2} \right) - \frac{3l}{m r^4}G^{2, 0, 0, 1}, \\
  \fl \dot{G}^{0,0,2,0} = -\frac{2}{mr^2} G^{0,0,1,1}  + \frac{4}{m r^{3}} \left(lG^{1, 0, 1, 0} + G^{1, 0, 1, 1} \right) - \frac{6l}{m r^4}G^{2, 0, 1, 0}, \\
  \fl \dot{G}^{1,0,1,0} = -\frac{1}{m} G^{0, 1, 1, 0} - \frac{1}{mr^2} G^{1, 0, 0, 1} + \frac{2}{m r^{3}} \left(lG^{2, 0, 0, 0} +G^{2, 0, 0, 1} \right) - \frac{3l}{m r^4}G^{3, 0, 0, 0}, \\
  \fl \dot{G}^{0,1,0,1} =  \left[\frac{3l^2}{2mr} - k
  \right]\frac{2}{r^{3}} G^{1, 0, 0, 1} - \left[\frac{2l^2}{mr} - k
  \right]\frac{2}{r^{4}} G^{2, 0, 0, 1} - \frac{1}{m r^{3}}\left(2l\Delta l^2 -G^{0, 0, 0, 3} \right) \nonumber \\
   + \frac{6l}{m r^{4}}G^{1, 0, 0, 2}, \\
  \fl \dot{G}^{0,1,1,0} = -\frac{1}{mr^2}  G^{0, 1, 0, 1} + \left[\frac{3l^2}{2mr} - k
  \right]\frac{2}{r^{3}} G^{1, 0, 1, 0} + \frac{1}{m r^{3}}  \left[2l\left(G^{1,1,0,0}-G^{0,0,1,1}\right)  \right]
  \nonumber \\
   -  \left(G^{0,0,1,2}-2G^{1,1,0,1}\right)- \left[\frac{2l^2}{mr} - k  \right]\frac{3}{r^{3}} G^{2, 0, 1, 0}
  + \frac{3l}{m r^{4}}  \left(2G^{1,0,1,1}-G^{2,1,0,0}\right).
\end{eqnarray}
We omit equations for $r$, $p_{\theta}$, $G^{2,0,0,0}$ and
$G^{1,0,0,1}$ because they are not modified with respect to previous
expressions of second order. Also we replace $G^{0,0,0,2}$ with
$\Delta l^2$.

Beside this system there are another eighteen equations:

\begin{eqnarray}
  \fl \dot{G}^{1,1,1,0} = -\frac{G^{0,2,1,0}}{m}-\frac{G^{1,1,0,1}}{m r^2}+\frac{3}{r^4} \left[-k+\frac{2 l^2}{m r}\right] G^{1,0,1,0} G^{2,0,0,0} + \frac{2}{r^3} \left[-k+\frac{3 l^2}{2 m
  r}\right]G^{2,0,1,0} \nonumber \\
   +\frac{3}{m r^4} \left[l (-2 G^{1,0,0,1} G^{1,0,1,0}+G^{1,1,0,0} G^{2,0,0,0}) - ( G^{1,0,0,2} G^{1,0,1,0} - G^{1,1,0,0} G^{2,0,0,1})\right]\nonumber \\
   -\frac{1}{m r^3}2 \left[\frac{1}{2} (-\Delta l^2 G^{1,0,1,0}+2 G^{1,0,0,1} G^{1,1,0,0})+l (G^{1,0,1,1}-G^{2,1,0,0})\right], \\
  \fl \dot{G}^{1,1,0,1} = -\frac{G^{0,2,0,1}}{m} + \frac{ \left( \Delta l^2 G^{1,0,0,1}-2l G^{1,0,0,2}\right)}{m r^3} - \frac{3 \left(2 l (G^{1,0,0,1})^2+G^{1,0,0,1} G^{1,0,0,2}\right)}{m r^4} \nonumber \\
   +\frac{3}{r^4} \left(-k+\frac{2 l^2}{m r}\right) G^{1,0,0,1} G^{2,0,0,0}+\frac{2}{r^3} \left(-k+\frac{3 l^2}{2 m r}\right) G^{2,0,0,1}, \\
  \fl \dot{G}^{1,0,1,1} = -\frac{G^{0,1,1,1}}{m}-\frac{G^{1,0,0,2}}{m r^2}+\frac{3 (l G^{1,0,0,1} G^{2,0,0,0}+G^{1,0,0,1} G^{2,0,0,1})}{m r^4}\nonumber \\
   -\frac{2 \left((G^{1,0,0,1})^2 - l G^{2,0,0,1}\right)}{m r^3}, \\
  \fl \dot{G}^{0,1,1,1} = -\frac{G^{0,1,0,2}}{m r^2}+\frac{2}{r^3} \left[-k+\frac{3 l^2}{2 m r}\right] G^{1,0,1,1} +\frac{3}{r^4} \left[-k+\frac{2 l^2}{m r}\right] G^{0,0,1,1} G^{2,0,0,0} \nonumber \\
   +\frac{3}{m r^4} \left[l (-2 G^{0,0,1,1} G^{1,0,0,1}+G^{0,1,0,1} G^{2,0,0,0})- ( G^{0,0,1,1} G^{1,0,0,2}- G^{0,1,0,1} G^{2,0,0,1})\right]  \nonumber \\
   -\frac{1}{m r^3} \left[ (-\Delta l^2 G^{0,0,1,1}+2 G^{0,1,0,1} G^{1,0,0,1})+2l (G^{0,0,1,2}-G^{1,1,0,1})\right],\\
  \fl \dot{G}^{1,2,0,0} = -\frac{G^{0,3,0,0}}{m} - \frac{3 (4 l G^{1,0,0,1} G^{1,1,0,0}+2 G^{1,0,0,2} G^{1,1,0,0})}{m r^4}+\frac{2 (\Delta l^2 G^{1,1,0,0}-2 l G^{1,1,0,1})}{m r^3}\nonumber\\
   +\frac{6}{r^4} \left[-k+\frac{2 l^2}{m r}\right] G^{1,1,0,0} G^{2,0,0,0}+\frac{4}{r^3} \left[-k+\frac{3 l^2}{2 m r}\right] G^{2,1,0,0}, \\
  \fl \dot{G}^{1,0,2,0} = -\frac{G^{0,1,2,0}}{m}+\frac{6 G^{1,0,1,0} (l G^{2,0,0,0}+ G^{2,0,0,1})}{m r^4}-\frac{4 ( G^{1,0,0,1} G^{1,0,1,0}- l G^{2,0,1,0})}{m r^3}, \\
  \fl \dot{G}^{1,0,0,2} = -\frac{G^{0,1,0,2}}{m}, \\
  \fl \dot{G}^{0,1,0,2} = \frac{\left( (\Delta l^2)^2-2l G^{0,0,0,3}\right)}{m r^3}+\frac{2}{r^3} \left[-k+\frac{3 l^2}{2 m r}\right] G^{1,0,0,2}+\frac{3}{r^4} \left[-k+\frac{2 l^2}{m r}\right] \Delta l^2 G^{2,0,0,0} \nonumber\\
   -\frac{3 \Delta l^2 (2 l G^{1,0,0,1}+ G^{1,0,0,2})}{m r^4},\nonumber\\%
  \fl \dot{G}^{0,1,2,0} = -\frac{2 G^{0,1,1,1}}{m r^2}+\frac{2}{r^3} \left[-k+\frac{3 l^2}{2 m r}\right] G^{1,0,2,0}+\frac{3}{r^4} \left[-k+\frac{2 l^2}{m r}\right] G^{0,0,2,0} G^{2,0,0,0} \nonumber\\
   -\frac{2}{m r^3} \Biggl[\left( 2 G^{0,1,1,0} G^{1,0,0,1}  -\Delta l^2 G^{0,0,2,0} + \frac{i \hbar}{2}\{ G^{1,1,0,0}-2G^{0,0,1,1} \}   \right) \Biggr. \nonumber\\
   \Biggl. + l (G^{0,0,2,1}-2 G^{1,1,1,0})\Biggr] +\frac{3}{m r^4} \Biggl[ \Biggl(2 G^{0,1,1,0} G^{2,0,0,1} - G^{0,0,2,0} G^{1,0,0,2} \Biggr. \Biggr. \nonumber \\
   \Biggl. \Biggl. +\frac{i \hbar}{2}\{ G^{2,1,0,0} -4 G^{1,0,1,1} \}\Biggr) + 2l ( G^{0,1,1,0} G^{2,0,0,0}- G^{0,0,2,0} G^{1,0,0,1})
   \Biggr],\\
  \fl \dot{G}^{0,2,0,1} = \frac{2 (\Delta l^2 G^{0,1,0,1}-2 l G^{0,1,0,2})}{m r^3}-\frac{6 (2 l G^{0,1,0,1} G^{1,0,0,1}+ G^{0,1,0,1} G^{1,0,0,2})}{m r^4}\nonumber \\
   +\frac{4}{r^3} \left[-k+\frac{3 l^2}{2 m r}\right] G^{1,1,0,1}+\frac{6}{r^4} \left[-k+\frac{2 l^2}{m r}\right] G^{0,1,0,1} G^{2,0,0,0}, \\
  \fl \dot{G}^{0,2,1,0} = -\frac{G^{0,2,0,1}}{m r^2}-\frac{4}{r^3}\left[k-\frac{3 l^2}{2 m r}\right] (G^{1,1,1,0}- G^{0,1,1,0} G^{1,0,0,0} )-\frac{6}{r^4} \left[k-\frac{2 l^2}{m r}\right] G^{0,1,1,0} G^{2,0,0,0} \nonumber\\
   -\frac{2}{m r^3} \left[ (G^{0,2,0,0} G^{1,0,0,1}- \Delta l^2 G^{0,1,1,0} )+l (2 G^{0,1,1,1}-G^{1,2,0,0})\right] \nonumber\\
   +\frac{3}{m r^4} \Biggl[l ( G^{0,2,0,0} G^{2,0,0,0} -4 G^{0,1,1,0} G^{1,0,0,1} + i \hbar \{ G^{0,0,1,1}-2  G^{1,1,0,0}\})\Biggr. \nonumber\\
   \Biggl. + \left( G^{0,2,0,0} G^{2,0,0,1} -2 G^{0,1,1,0} G^{1,0,0,2} + \frac{i \hbar }{2}\{ G^{0,0,1,2}-4  G^{1,1,0,1}\} \right)\Biggr], \\
  \fl \dot{G}^{2,0,0,1} = -\frac{2 G^{1,1,0,1}}{m}, \\
  \fl \dot{G}^{2,0,1,0} = -\frac{2 G^{1,1,1,0}}{m}-\frac{G^{2,0,0,1}}{m r^2}+\frac{3 \left(l (G^{2,0,0,0})^2+G^{2,0,0,0} G^{2,0,0,1}\right)}{m r^4}-\frac{2 (G^{1,0,0,1} G^{2,0,0,0}-l G^{3,0,0,0})}{m r^3}, \\
  \fl \dot{G}^{2,1,0,0} = -\frac{2 G^{1,2,0,0}}{m}+\frac{3}{r^4} \left[-k+\frac{2 l^2}{m r}\right] (G^{2,0,0,0})^2 +\frac{2}{r^3} \left[-k+\frac{3 l^2}{2 m r}\right] G^{3,0,0,0} \nonumber\\
   + \frac{\left( \Delta l^2 G^{2,0,0,0}-2l G^{2,0,0,1}\right)}{m r^3} - \frac{3 (2 l G^{1,0,0,1} G^{2,0,0,0}+G^{1,0,0,2} G^{2,0,0,0})}{m r^4}, \\
  \fl \dot{G}^{0,0,0,3} = 0, \label{0003}\\
  \fl \dot{G}^{0,0,3,0} = -\frac{3 G^{0,0,2,1}}{m r^2}-\frac{6 ( G^{0,0,2,0} G^{1,0,0,1}- l G^{1,0,2,0})}{m r^3}+\frac{9 ( l G^{0,0,2,0} G^{2,0,0,0}+ G^{0,0,2,0} G^{2,0,0,1})}{m r^4}, \\
  \fl \dot{G}^{0,3,0,0} = \frac{\left(3 \Delta l^2 G^{0,2,0,0}-6 l G^{0,2,0,1}\right)}{m r^3}+\frac{6}{r^3} \left[-k+\frac{3 l^2}{2 m r}\right] G^{1,2,0,0}  \nonumber\\
   -\frac{9 (2 l G^{0,2,0,0} G^{1,0,0,1}+ G^{0,2,0,0} G^{1,0,0,2})}{m r^4} + \frac{9}{r^4}\left[k - \frac{2 l^2}{m r}\right] ( i \hbar G^{1,1,0,0}- G^{0,2,0,0} G^{2,0,0,0}), \\
  \fl \dot{G}^{3,0,0,0} = -\frac{3 G^{2,1,0,0}}{m}.
\end{eqnarray}

From (\ref{0003}) we immediately see that $G^{0,0,0,3}$ is a
constant, as we already noticed for the moments of $G^{0,0,0,n}$ to
any order $n$.


\section*{References}
\begin{thebibliography}{99}
\bibitem{Q} E. Merzbacher, \emph{Quantum Mechanics}, 2nd edition Wiley, New York,
(1970).

\bibitem{Eff} M. Bojowald and A. Skirzewski, Rev. Math. Phys.
\textbf{18}, 713 (2006).


\bibitem{EffAcc} F. Cametti, G. Jona-Lasinio, C. Presilla and F. Toninelli In Proceedings of the International School of Physics ``Enrico Fermi'', Course CXLIII, Ed. G. Casati, I. Guarneri, U. Smilansky (IOS Press, Amsterdam 2000), p. 431-448 (quant-ph/9910065); W. Heisenberg, H. Euler, Zeitschr. Phys. \textbf{98}, 714 (1936), (physics/0605038).
\bibitem{Dias} N. C. Dias, A. Mikovic, and J. N. Prata, J. Math. Phys.
(N.Y.) \textbf{47}, 082101 (2006).


\bibitem{MB} M. Bojowald, Phys. Rev. D \textbf{75}, 081301(R) (2007); M. Bojowald, Phys. Rev. D \textbf{75}, 123512 (2007).
\bibitem{MBNat} M. Bojowald, Nature Physics \textbf{3}, 523 (2007).
\bibitem{MB-H} M. Bojowald, H. H. Hern\'{a}ndez and A. Skirzewski, Phys. Rev. D \textbf{76}, 063511 (2007).

\bibitem{D} M. Bojowald, D. Brizuela, H. H. Hern\'{a}ndez, M. J. Koop and H. A. Morales-T\'{e}cotl,
Phys. Rev. D \textbf{84}, 043514 (2011).

\bibitem{H1} M. Bojowald, M. Kagan, H. H. Hern\'{a}ndez, A. Skirzewski,
Phys. Rev. D \textbf{75}, 064022 (2007); M. Bojowald and A.
Tsobanjan, Phys. Rev. D \textbf{80}, 125008 (2009).

\bibitem{LQC-dy} M. Bojowald, arXiv:1101.5592, (2011).

\bibitem{Dirac} P.A.M. Dirac, \emph{The Principles of Quantum Mechanics}, First edition, Oxford Clarendon Press, (1930).

\bibitem{1DH} R. Loudon, Am. J. Phys. \textbf{27}, 649 (1959); H. N.
N\'{u}\~{n}ez-Y\'{e}pez, A. L. Salas-Brito, D. A. Solis, Phys. Rev.
A \textbf{83}, 064101 (2011).

\bibitem{1DHaps} M. Mayle, B. Hezel, I. Lesanovsky and P. Schmelcher, Phys. Rev.
Lett. \textbf{99}, 113004 (2007); M. M. Nieto, Phys. Rev. A
\textbf{61}, 034901 (2000); C. D. Schwieters and J. B. Delos, Phys.
Rev. A \textbf{51}, 1030 (1995).


\bibitem{kosteleky} R. Bluhm and V. A. Kosteleck\'{y}, Phys. Rev. A \textbf{48}, R4047
(1993); R. Bluhm, V. A. Kosteleck\'{y} and B. Tudose, Phys. Rev. A
\textbf{52}, 2234 (1995).

\bibitem{wkb} R. Langer, Phys. Rev. {\bf 59} 669, (1937).

\bibitem{brown} L. S. Brown, Am. J. Phys. \textbf{41}, 525 (1973);

\bibitem{cole} D. C. Cole and Yi Zou, Phys. Rev. E \textbf{69},
016601 (2004).

\bibitem{stroud} Z. D. Gaeta and C. S. Stroud Jr., Phys. Rev. A \textbf{42}, 6308
(1990).

\bibitem{nieto} M. M. Nieto, Phys. Rev. D \textbf{22}, 391 (1980).

\bibitem{Ghosh} D. Bhaumik, B. D. Dutta-Roy and G. Ghosh, J. Phys.
A: Math. Gen. \textbf{19}, 1355 (1986); S. Nandi and C. S. Shastry,
J. Phys. A: Math. Gen. \textbf{22}, 1005 (1989); A. Rauh and J.
Parisi, Phys. Rev. A \textbf{83}, 042101 (2011).

\bibitem{exps1} J. Parker and C. S. Stroud Jr., Phys. Rev. Lett. \textbf{56}, 716
(1986); J. Parker and C. S. Stroud, Phys. Scripta T12, 70 (1986); J.
A. Yeazell and C. S. Stroud Jr., Phys. Rev. Lett. \textbf{60}, 1494
(1988); J. A. Yeazell and C. S. Stroud Jr., Phys. Rev. A
\textbf{43}, 5153 (1991).

\bibitem{exps2}  J. J. Mestayer, B. Wyker, J. C. Lancaster, F. B. Dunning, C. O. Reinhold, S. Yoshida, and J.
Burgd\"{o}rfer, Phys. Rev. Lett. \textbf{100}, 243004 (2008); H.
Maeda, J. H. Gurian, and T. F. Gallagher, Phys. Rev. Lett.
\textbf{102}, 103001 (2009); J. J. Mestayer, B. Wyker, F. B.
Dunning, S. Yoshida, C. O. Reinhold, and J. Burgd\"{o}rfer Phys.
Rev. Lett. \textbf{79}, 033417 (2010).

\end{thebibliography}
\end{document}